\documentclass[twocolumn,showpacs,preprintnumbers,amsmath,amssymb,epsfig,widetext]{revtex4}

\usepackage{graphicx}
\usepackage{dcolumn}
\usepackage{bm}
\usepackage{epsfig}

\usepackage{float}
\usepackage{amsmath}
\usepackage{epsfig,floatflt}
\usepackage{subfigure}

\newcommand{\la}{\lesssim}
\newcommand{\ga}{\gtrsim}

\def\fdg{\hbox{$.\!\!^\circ$}}

\begin{document}


\title{The Spatially Closed Universe}

\author{Chan-Gyung Park}
\email{park.chan.gyung@gmail.com}
\affiliation{Astrophysical Research Center for the Structure and Evolution
             of the Cosmos, Sejong University, Seoul, 143-747, Korea}

\date{\today}


\begin{abstract}
The general world model for homogeneous and isotropic universe has been
proposed. For this purpose, we introduce a global and fiducial system
of reference (world reference frame) constructed on a $5$-dimensional
space-time that is embedding the universe, and define the line element
as the separation between two neighboring events that are distinct in space
and time, as viewed in the world reference frame.
The effect of cosmic expansion on the measurement of physical distance
has been correctly included in the new metric, which differs from
the Friedmann-Robertson-Walker metric where the spatial separation
is measured for events on the hypersurface at a constant time while
the temporal separation is measured for events at different time epochs.
The Einstein's field equations with the new metric imply that
closed, flat, and open universes are filled with positive, zero,
and negative energy, respectively.
The curvature of the universe is determined by the sign of mean energy density.
We have demonstrated that the flat universe is empty and stationary,
equivalent to the Minkowski space-time,
and that the universe with positive energy density is always spatially
closed and finite.
In the closed universe, the proper time of a comoving observer does not
elapse uniformly as judged in the world reference frame, in which both
cosmic expansion and time-varying light speeds cannot exceed the limiting
speed of the special relativity.
We have also reconstructed cosmic evolution histories of the closed
world models that are consistent with recent astronomical observations,
and derived useful formulas such as energy-momentum relation of particles,
redshift, total energy in the universe, cosmic distance and time scales,
and so forth. It has also been shown that the inflation
with positive acceleration at the earliest epoch is improbable.
\end{abstract}


\maketitle


\section{Introduction}
\label{sec:intro}

The main goal of modern cosmology is to build a cosmological model
that is consistent with astronomical observations. To achieve this
goal, tremendous efforts have been made both on theories and on observations
since the general theory of relativity was developed.
So far the most successful model of the universe is the 
Friedmann-Robertson-Walker (FRW) world model \cite{fri22,fri24,rob29,wal35}.
The FRW world model predicts reasonably well the current observations
of the cosmic microwave background (CMB) radiation and the large-scale
structures in the universe.
The precisely determined cosmological parameters of the FRW world model 
imply that our universe is consistent with the spatially flat world model 
dominated by dark energy and cold dark matter ($\Lambda\textrm{CDM}$)
with adiabatic initial condition driven by inflation
\cite{spergel07,tegmark06}. 

Although the flat FRW world model is currently the most reliable
physical world model, one may have the following fundamental questions
on the nature of the FRW world model.
First, mathematically, if a space-time manifold is flat, then the Riemann
curvature tensor should vanish, and vice versa.
However, the Riemann curvature tensor of the flat FRW world model does
not vanish unless the cosmic expansion speed and acceleration are zeros,
which implies that the physical space-time of the flat FRW world is not
geometrically flat but curved. Only its spatial section at a constant time
is flat.

Secondly, the cosmic evolution equations of the FRW world model 
can be derived from an application of the Newton's gravitation and
the local energy conservation laws to the dynamical motion of an expanding
sphere with finite mass density \cite{milne34,mccrea34}. 
Besides, the Newton's gravitation theory has been widely
used to mimic the non-linear clustering of large-scale structures 
in the universe even on the horizon-sized $N$-body simulations
\cite{colberg2000,park05}.
On large scales, the close connection between the FRW world model
and the Newton's gravitation law is usually attributed to the fact
that the linear evolution of large-scale density perturbations satisfies
the weak gravitational field condition.
Recently, Hwang and Noh \cite{hwang06} show that the relativistic fluid
equations perturbed to second order in a flat FRW background world coincide
exactly with the Newtonian results, and prove that the Newtonian numerical
simulation is valid in all cosmological scales up to the second order.
However, one may have a different point of view that the Newton's
gravitational action at a distance appears to be valid even on the
super-horizon scales in the FRW world just because the world model
does not reflect the full nature of the relativistic theory of gravitation.

Thirdly, according to the FRW world model, the universe at sufficiently 
early epoch ($z \gtrsim 1000$) is usually regarded as flat since the curvature
parameter contributes negligibly to the total density. The present non-flat
universe should have had the density parameter approaching to $\Omega = 1$
with infinitely high precision just after the big-bang (flatness problem).
On the other hand, if we imagine the surface of an expanding balloon
with positive curvature, then the curvature of the surface is always
positive and becomes even higher as the balloon is traced back to
the earlier epoch when it was smaller.
This prediction from the common sense contradicts the FRW world model.

Observationally, the flat $\Lambda\textrm{CDM}$ universe is favored
by the recent joint cosmological parameter estimation using the Wilkinson
Microwave Anisotropy (WMAP) CMB \cite{hinshaw07,page07}, large-scale
structures \cite{cole05,tegmark04}, type Ia supernovae (SNIa;
\cite{riess07,wood07}), Hubble constant \cite{freedman01,macri06,sandage06},
baryonic oscillation data \cite{eisen05}, and so on.
However, the WMAP CMB data alone is more compatible with the non-flat
FRW world model ($\S7.3$ of \cite{spergel07} and Table III
of \cite{tegmark06}).
Besides, some parameter estimations using SNIa data or angular size-redshift
data of distant radio sources alone suggest a possibility of the closed
universe \cite{clocchi06,jackson06}.
The combinations of the WMAP plus the SNIa data or the Hubble constant data
also imply the possibility of the closed universe, giving curvature
parameters $\Omega_k = -0.011\pm 0.012$ and $\Omega_k = -0.014\pm 0.017$,
respectively \cite{spergel07}, although the estimated values are still
consistent with the flat FRW world model.

The questions above and the observational constraints on the cosmological
model may bring about possibilities of non-flat or non-FRW world models.
Interestingly, Einstein claimed that our universe is spatially bounded
or closed \cite{ein22}. The primary reason for his preference to the
closed universe is because Mach's idea \cite{mach93,misner73} that
the inertia depends upon the mutual action of bodies is compatible only
with the finite universe, not with a quasi-Euclidean, infinite universe.
According to Einstein's argument, an infinite universe is possible only
if the mean density of matter in the universe vanishes, which is unlikely
due to the fact that there is a positive mean density of matter in the
universe \footnote{However, in the appendix to the second edition of
his book \cite{ein22}, Einstein summarized Friedmann's world models
and discussed a universe with vanishing spatial curvature and non-vanishing
mean matter density, which differs from his original argument.}. 

In this paper, we propose the general world model for homogeneous and
isotropic universe which supports Einstein's perspective on the physical
universe. The outline of this paper is as follows.
In Sec. \ref{sec:metric}, we consider the effect of cosmic expansion 
on the physical space-time distance between neighboring events and
describe how to define the line element for homogeneous and isotropic
universes of various spatial curvature types.
The metric and the cosmic evolution equations for flat, closed, and open
universes are derived in Sec. \ref{sec:nspace}. It will be shown that our
universe is spatially closed.  
In Sec. \ref{sec:some}, we reconstruct cosmic evolution histories of the 
closed world models, and derive interesting properties of the closed
universe. In Sec. \ref{sec:inflation}, we discuss whether the inflation
theory is compatible with the closed world model or not.
Conclusion follows in Sec. \ref{sec:conc}.

Throughout this paper, we adopt a sign convention $(+,-,-,-)$ 
for the metric tensor $g_{ik}$, and denote a 4-vector in space-time as 
$p^{i}$ ($i=0,1,2,3$) and a 3-vector in space as $p^{\alpha}$
($\alpha=1,2,3$) or $\mathbf{p}$. The Einstein's field equations are
\begin{equation}
  R_{ik} - \frac{1}{2} g_{ik}R = 8\pi G T_{ik}+\Lambda g_{ik},
\label{eq:einstein}
\end{equation}
where $R_{ik}=R^{a}_{~iak}$ is the Ricci tensor, $R=R^i_{~i}$ the Ricci
scalar, $T_{ik}$ the energy-momentum tensor, $G$ the Newton's gravitational
constant, and $\Lambda$ the cosmological constant.
The Riemann curvature tensor is given by
$R^{a}_{~ibk}
    =\partial_b \Gamma^{a}_{ki}-\partial_{k}\Gamma^{a}_{bi}
    +\Gamma^{a}_{bn} \Gamma^{n}_{ki}-\Gamma^{a}_{kn}\Gamma^{n}_{bi}$,
with the Christoffel symbol 
$\Gamma^{a}_{ik}={1\over 2} g^{ab} 
(\partial_i g_{kb}+\partial_k g_{ib} -\partial_b g_{ik})$. 
The energy-momentum tensor for perfect fluid is
\begin{equation}
   T_{ik} = (\varepsilon_\textrm{b} + P_\textrm{b})u_i u_k
          - P_\textrm{b} g_{ik},
\label{eq:emtensor}
\end{equation}
where $\varepsilon_\textrm{b}$ and $P_\textrm{b}$ are background energy density
and pressure of ordinary matter and radiation, and $u_{i}$ is the 4-velocity
of a fundamental observer.
We assume that the cosmological constant acts like a fluid
with effective energy density $\varepsilon_\Lambda = \Lambda/8\pi G$
and pressure $P_\Lambda = -\varepsilon_\Lambda$.
The limiting speed in the special theory of relativity is set to unity 
($c\equiv 1$).

\section{How to define metric for homogeneous and isotropic universes?}
\label{sec:metric}

The starting point for constructing a physical world model is to define 
the space-time separation between two neighboring events,
i.e., the line element
\begin{equation}
   d s^2 = g_{ik}(x) d x^i d x^k,
\label{eq:ds2}
\end{equation}
where $g_{ik}(x)$ is the metric tensor which determines all the geometric
properties of space-time in a system of coordinates.
In (\ref{eq:ds2}), the two events are generally distinct in space and time,
separated by $dx^i=(dx^0, dx^1, dx^2, dx^3)$.
In the special theory of relativity, a separation between two distinct events
is given by
\begin{equation}
   d s^2 = d t^2 - d\mathbf{r}^2,
\label{eq:metric_sr}
\end{equation}
which is invariant in all inertial reference frames.
The metric $g_{ik}=\textrm{diag}(1,-1,-1,-1)$ is called
the Minkowski metric.

The early development in modern cosmology was focused on finding
the metric appropriate for the real universe whose space-time structure
is inconsistent with the Minkowski metric due to the expansion
of the universe and the presence of matter in it.
As a pioneer, Einstein \cite{ein17} developed a model of the static closed
universe that is spatially homogeneous and isotropic,
by adopting a metric with $g_{00}=1$ and $g_{0\alpha}=0$
from the static condition.
Friedmann \cite{fri22} developed a more general world model that includes
both stationary and non-stationary universes of positive spatial curvature,
by assuming that one can make $g_{00}=1$ and $g_{0\alpha}=0$
by an appropriate choice of time coordinate. 
Weyl \cite{weyl23} postulated that in a cosmological model the world lines
of particles (e.g., galaxies) form a $3$-bundle of nonintersecting geodesics
orthogonal to a series of space-like hypersurfaces, which implies that
$g_{00}$ depends on $x^0 = t$ only and $g_{0\alpha}=0$.
Robertson \cite{rob29} also argued that the line element may be expressed as
$d s^2=d t^2 + g_{\alpha\beta} d x^\alpha d x^\beta$ for a universe 
where the matter has on the whole the time-like geodesics 
$x^\alpha =\textrm{const.}$ as world lines, and the coordinate $t$
can be interpreted as a mean time which serves to define
proper time and simultaneity for the entire universe.
Walker \cite{wal35,wal37} demonstrated that the line element for
the 3-dimensional Riemannian space of constant curvature,
$d\sigma^2 = h_{\alpha\beta}d x^\alpha d x^\beta$,
is invariant under all transformations belonging to a group of motions $G_6$,
and proposed a metric for the non-stationary space-time manifold as
$d s^2=d t^2-a^2(t) d\sigma^2$, where $t$ is a physical time of
an observer and is invariant under a transformation from one observer
to another. 

The resulting line element for spatially homogeneous and isotropic universe 
(FRW metric) is concisely written as
\begin{equation}
\begin{split}
   d s^2 &= dt^2 -a^2 (t) \mskip+6mu d\sigma^2  \\
         &= d t^2 - a^2(t)\left[ d\chi^2 + S_k^2(\chi)( d\theta^2 
             + \sin^2\theta d\phi^2) \right],
\end{split}
\label{eq:FRW_metric}
\end{equation}
where $a(t)$ is a cosmic expansion scale factor,
$d\sigma$ is the comoving-space separation between events,
and $S_k(\chi) = \sinh\chi$, $\chi$, and $\sin\chi$ for open ($k=-1$), 
flat ($k=0$), and closed ($k=+1$) spaces, respectively
(see Ref. \cite{weinberg72} for a detail derivation of the FRW metric).
The corresponding cosmic evolution equations known as Friedmann equations are
\begin{equation}
   {{\ddot{a}}\over{a}} = -{{4\pi G}\over{3}}(\varepsilon_\textrm{b}
             +3P_\textrm{b}) + {{\Lambda}\over{3}},
\label{eq:frw1}
\end{equation}
and
\begin{equation}
   \left({{\dot{a}}\over{a}}\right)^2 
       = {{8\pi G}\over{3}}\varepsilon_\textrm{b}
        -{{k}\over{a^2}} +{{\Lambda}\over{3}}.
\label{eq:frw2}
\end{equation}
The dot denotes a differentiation with respect to time.  

We note that previous studies all assumed that $g_{00}=1$ and $g_{0\alpha}=0$ 
in the form of line element, where spatial and temporal distances
were considered to be separate.
In particular, a serious inconsistency is seen in the FRW metric.
The real-space separation $a(t)d\sigma$ is measured for events
on the hypersurface at a constant time (a slice of simultaneity).
On the other hand, the temporal separation $dt$ is measured for events
at different time epochs.
In other words, spatial and temporal distances are related
to two different couples of events, in contradiction to the general
definition of the line element (\ref{eq:ds2}) in which only a couple of events
should be involved.

Besides, in the FRW metric the geometry of physical space separates
into the geometry of comoving space and the expansion history $a(t)$.
This is the reason why the flat FRW world model is geometrically flat
only in the comoving space, not in the physical space and time.
Therefore, the effect of cosmic expansion on the space-time
geometry has not been correctly reflected in the FRW metric,
giving an inaccurate measure of distance between events.

To derive the general form of line element for homogeneous
and isotropic universe, let us introduce a $4+1$ Minkowski space-time
composed of 4-dimensional Euclidean space and 1-dimensional time,
and assume that our universe is spatially a 3-dimensional hypersurface
with uniform curvature embedded in the 4-dimensional space.
In fact, the hypersurface with negative curvature (open space) 
cannot be embedded in the Euclidean space.
Here we restrict our attention to flat and closed spaces,
deferring the discussion about the open space to Sec. \ref{sec:open}.

Throughout this paper, we call a reference frame constructed on
the high dimensional Minkowski space-time as {\it world reference frame}.
It is a global system of reference provided with a rigid measuring rod
and a number of clocks to indicate position and time ({\it world time})
of an event. Distances between events on the universe will be measured
based on this fiducial system.

Although the world reference frame has been introduced for mathematical
convenience, it is useful in that the space-time coordinates
of the frame are independent of the dynamics of the universe. 
As will be shown later, the proper time as measured by an observer
in the universe does not elapse uniformly, being affected by the cosmic
expansion. Thus, it is natural to use the world time coordinate
with uniform lapse for a fair comparison of physical phenomena
in the expanding universe.
The space coordinates of the world reference frame can be conveniently 
used to describe the geometry of a 3-dimensional hypersurface embedded in the
4-dimensional Euclidean space (see Sec. \ref{sec:closed}).

\begin{figure}
\mbox{\epsfig{file=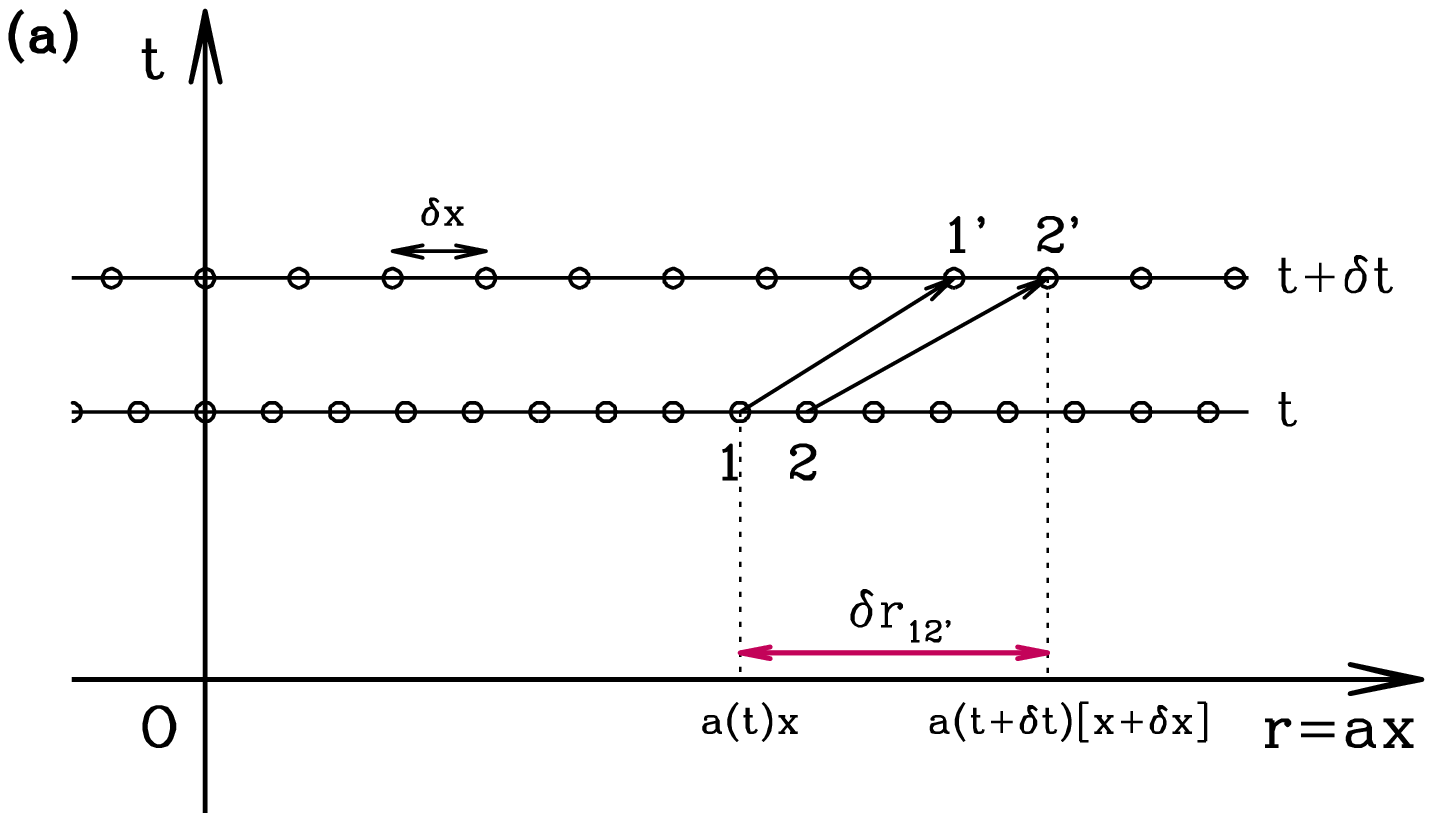,width=88mm,clip=}}
\mbox{\epsfig{file=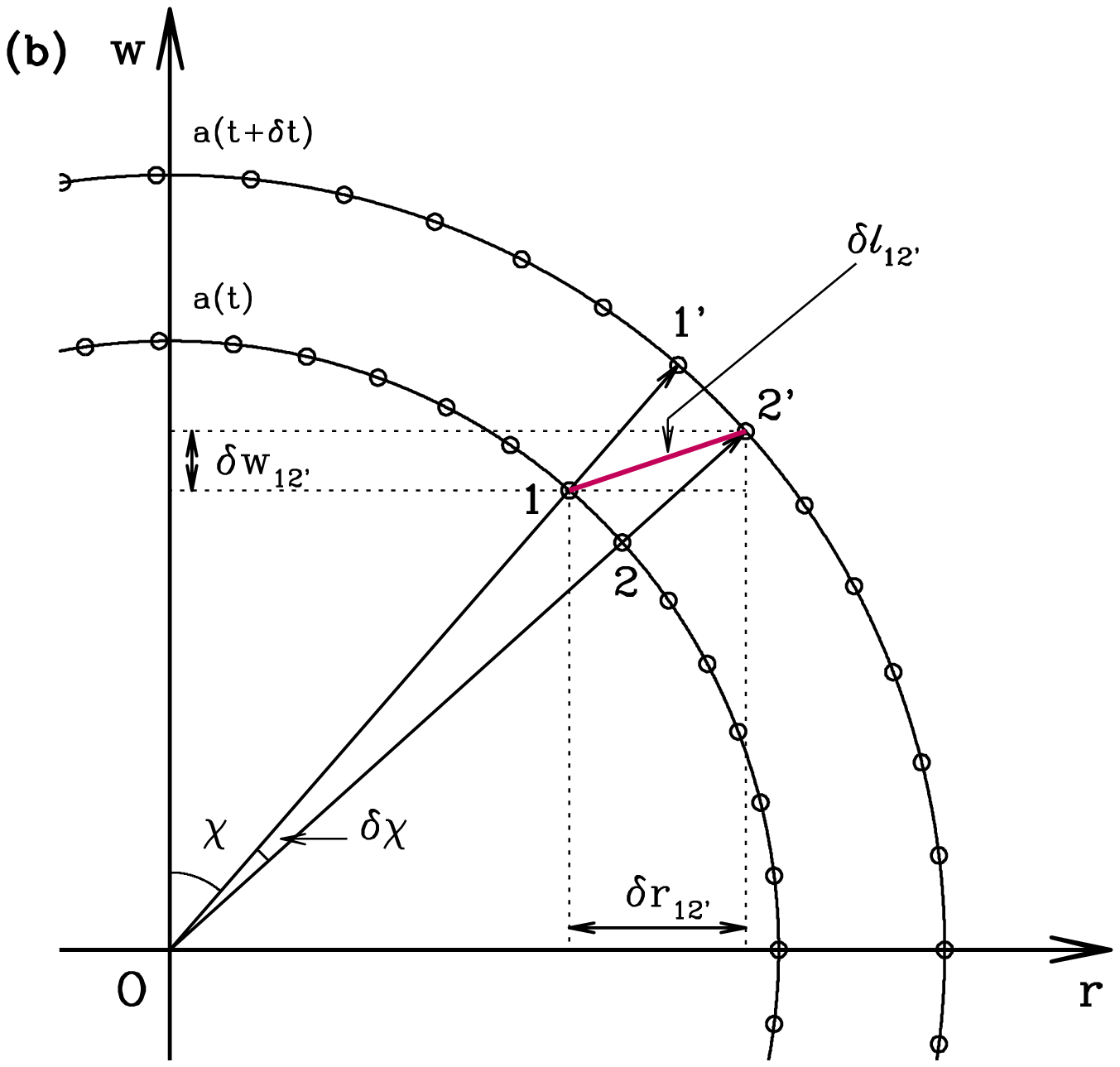,width=88mm,clip=}}
\caption{Schematic diagrams showing the expansion of (a) flat and
(b) closed 1-dimensional homogeneous and isotropic hypersurfaces.
Events denoted as open circles are equally spaced out on two
hypersurfaces that are infinitesimally separated by $\delta t$.
Note that the time axis, which is orthogonal to $r$- and $w$-axes,
is omitted in (b). The line element is defined as the space-time
separation between two distinct events $1$ and $2'$ (see text).}
\label{fig:line}
\end{figure}

An example of the expanding 1-dimensional flat hypersurface
(with uniform and zero curvature) is shown in Fig. \ref{fig:line}(a).
At initial time $t$, a hypersurface is given as a straight line,
on which there are equally spaced events (open circles) 
with mutual comoving separation $\delta x$, all at rest with respect
to the comoving coordinate system whose spatial coordinate $x$ is related
to the real-space coordinate by $r=a(t)x$.
After an infinitesimal time $\delta t$, the straight line has been
expanded, and the proper separation between neighboring events on the
hypersurface has increased from $a(t)\delta x$ to $a(t+\delta t) \delta x$. 

We define the line element $\delta s^2$ as the space-time separation
between two {\it distinct} events located at $(t,r)$ and
$(t+\delta t, r+\delta r)$, which correspond to events
$1$ and $2'$ in Fig. \ref{fig:line}(a) without loss of generality.  
The spatial separation between events $1$ and $2'$ as measured 
in the world reference frame is
$\delta r_{12'} = a(t+\delta t)[x+\delta x] - a(t)x \simeq
\dot{a}(t) x\delta t + a(t) \delta x$ up to the first order of
$\delta t$ and $\delta x$. Therefore, we get
\begin{equation}
   \delta s_{12'}^2 = \delta t^2 - \delta r_{12'}^2 
        = (1-\dot{a}^2 x^2)\delta t^2 - 2\dot{a} a x\delta t\delta x
              -a^2 \delta x^2,
\label{eq:lineflat}
\end{equation}
where the effect of cosmic expansion on the physical separation
between events has been included explicitly.
Note that the line element (\ref{eq:lineflat}) implies that generally
$g_{00}$ is a function of both time- and space-coordinates 
and $g_{0\alpha}\ne 0$, which violates the Weyl's postulate unless $\dot{a}=0$.

The line element for non-flat space can be defined analogously.
As an example of the closed space, Fig. \ref{fig:line}(b) shows 
an expanding 1-dimensional circle (1-sphere) with radius $a(t)$
at two distinct (infinitesimally separated) world times. 
The expanding circle is a $1$-dimensional hypersurface
with uniform and positive curvature embedded in the 2-dimensional
Euclidean space, $rw$-plane.
Events on the circle are equally spaced out with $\delta\chi$,
where $\chi$ is a comoving coordinate related to $r$-coordinate
by $r=a\sin\chi$. 
The line element is defined as the space-time separation between
distinct events $1$ and $2'$, written concisely as
\begin{equation}
   \delta s_{12'}^2 = \delta t^2 - \delta l_{12'}^2
      = (1-\dot{a}^2) \delta t^2 -a^2 \delta\chi^2,
\label{eq:lineclosed}
\end{equation}
where $\delta l_{12'}^2 = \delta r_{12'}^2 + \delta w_{12'}^2$
is the spatial separation between events $1$ and $2'$ as measured
in the world reference frame.
The spatial distances projected on $r$- and $w$-axes are given by
$\delta r_{12'} = a(t+\delta t)\sin(\chi+\delta \chi) - a(t) \sin\chi$ 
and $\delta w_{12'} = a(t+\delta t)\cos(\chi+\delta \chi)-a(t)\cos\chi$,
respectively. In the second equality of (\ref{eq:lineclosed}),
$\delta r_{12'}$ and $\delta w_{12'}$ have been expanded up to the first 
order of $\delta t$ and $\delta\chi$.
From two cases, it is clear that the expansion of space affects
both space and time intervals in the line element, which is the main
difference from the FRW metric.
Generally, the line element (metric) should reflect the fact that
the cosmic expansion is a dynamical phenomenon. 

Eq. (\ref{eq:lineflat}) implies that the FRW line element
for the 1-dimensional flat space is valid only at a local region
around an observer at $x=0$.
The FRW line element for the closed space also has a similar form
to the flat case, i.e., $\delta s^2 = \delta\tau^2 - a^2 \delta\chi^2$,
where, to be consistent with (\ref{eq:lineclosed}),
$\delta\tau$ should be interpreted as a proper time interval 
measured by a local observer ($\delta\tau = \delta t \sqrt{1-\dot{a}^2}$).
To such an observer who may be located between events $1$ and $2$
(or between events $1'$ and $2'$ after $\delta t$),
the spatial separation between neighboring events appears to be $a\delta\chi$
[e.g., the arc length between events 1 and 2 in Fig. \ref{fig:line}(b)].
Therefore, both FRW line elements describe the space-time separation
between events near an observer, which is the local nature of the FRW metric.

\section{Metric and evolution equations of expanding universes}
\label{sec:nspace}

In this section, we define the general forms of metric for homogeneous
and isotropic universes of various spatial curvature types, and derive
the corresponding cosmic evolution equations from the Einstein's field
equations.

\subsection{Flat universe}
\label{sec:flat}

Suppose that the universe is spatially an expanding 3-dimensional flat
hypersurface (with uniform and zero curvature) embedded in a 4-dimensional
Euclidean space with the Cartesian coordinates $(r_1, r_2, r_3, r_4)$.
The embedded flat space is infinite, homogeneous, and isotropic,
and is Euclidean at an instant of time.
To simplify the problem, let us assume that the hypersurface is orthogonal
to the $r_4$-axis so that the fourth Cartesian coordinate can be ignored.
Then, each event on the flat hypersurface is labelled by the world time $t$
and the real-space position $\mathbf{r}$ defined as 
\begin{equation}
   \mathbf{r} = a(t) \mathbf{x},
\label{eq:rax}
\end{equation}
where $\mathbf{x}$ is the comoving-space position vector
with the Cartesian coordinates
$(x_1,x_2,x_3)=(x\sin\theta\cos\phi, x\sin\theta\sin\phi, x\cos\theta)$
in relation to the comoving spherical coordinate
$x^\alpha = (x,\theta,\phi)$ with $x=|\mathbf{x}|=(x_1^2+x_2^2+x_3^2)^{1/2}$.

With the help of the differential of (\ref{eq:rax})
\begin{equation}
   d \mathbf{r} = \dot{a}(t) d t \mathbf{x}+a(t) d\mathbf{x},
\label{eq:diff_rax}
\end{equation}
the line element is defined as the space-time separation between events
located at $(t,\mathbf{r})$ and $(t+d t,\mathbf{r}+d\mathbf{r})$:
\begin{equation}
\begin{split}
   d s^2 &= d t^2 -d\mathbf{r}^2 
      =\left(1-\dot{a}^2 \mathbf{x}^2 \right)d t^2 
     -2\dot{a} a d t \mathbf{x}\cdot d\mathbf{x}
     -a^2 d\mathbf{x}^2 \\
     &= \left(1-\dot{a}^2 x^2 \right)d t^2
        - 2\dot{a} a x d t d x  \\
     &\mskip+100mu -a^2 \left[d x^2+x^2(d\theta^2+\sin^2\theta d\phi^2) \right].
\end{split}
\label{eq:line_flat}
\end{equation}
The metric tensor in the coordinate system $x^i=(t,x,\theta,\phi)$ is
\begin{equation}
g_{ik} = \left(
\begin{array}{cccc}
1-\dot{a}^2 x^2 & -\dot{a}ax & 0 & 0 \\
-\dot{a} ax & -a^2 & 0 & 0 \\
0 & 0 & -a^2 x^2 & 0 \\
0 & 0 & 0 & -a^2 x^2 \sin^2\theta 
\end{array}  \right).
\label{eq:metric_flat}
\end{equation}

We calculate the Christoffel symbols and the Riemann curvature tensor 
from the metric tensor. The non-zero Christoffel symbols  
$\Gamma^{a}_{ik}$ ($=\Gamma^{a}_{ki}$) are
\begin{equation}
\begin{split}
   \Gamma^{1}_{00} &= {{\ddot{a}}\over a}x, \mskip+12mu
      \Gamma^{1}_{01}=\Gamma^{2}_{02}=\Gamma^{3}_{03}={\dot{a} \over a}, \\
   \Gamma^{1}_{22} &= -x, \mskip+12mu
      \Gamma^{1}_{33} =-x\sin^2\theta, \mskip+12mu
      \Gamma^{2}_{12} = \Gamma^{3}_{13}={1 \over x}, \\
   \Gamma^{2}_{33} &=  -\sin\theta\cos\theta, \mskip+12mu
      \Gamma^{3}_{23}=\cot\theta.
\end{split}
\end{equation}
One easily verifies that the Riemann curvature tensor vanishes: 
\begin{equation}
R^{a}_{~ibk} = 0.
\label{eq:raibk}
\end{equation}
Thus the Ricci tensor $R_{ik}$ and the Ricci scalar $R$ also vanish.
This demonstrates explicitly that the space-time curvature of the expanding
flat universe is zero.

The proper time interval as measured by an observer in arbitrary motion
is obtained from (\ref{eq:line_flat}) as
\begin{equation}
   d\tau = d t \left[1-\dot{a}^2 x^2 -2\dot{a} a x
       \left(\frac{d x}{d t}\right) -a^2 v^2 \right]^{1/2},
\label{eq:flat_tau}
\end{equation}
where $a v = (-v_\alpha v^\alpha)^{1/2}$ is the magnitude of proper
3-velocity, $v^\alpha = d x^\alpha/d t$ is the 3-velocity
in the comoving coordinate system, and $v_\alpha = g_{\alpha\beta} v^\beta$.
For an observer who is at rest ($v^\alpha=0$) in the comoving coordinate
system (hereafter a comoving observer), we get the proper time interval
\begin{equation}
   d\tau = d t (1-\dot{a}^2 x^2)^{1/2}
\label{eq:flat_tauc}
\end{equation}
and the 4-velocity
\begin{equation}
   u^{i} = \frac{d x^i}{d\tau}
         = \left(\frac{1}{\sqrt{1-\dot{a}^2 x^2}},0,0,0\right)
\label{eq:ui_flat}
\end{equation}
of the observer. Inserting (\ref{eq:metric_flat}) and (\ref{eq:ui_flat})
into (\ref{eq:emtensor}) gives the energy-momentum tensor,
whose non-zero components are
\begin{equation}
\begin{split}
   T_{00} &= \varepsilon_\textrm{b} (1-\dot{a}^2 x^2), \mskip+12mu
      T_{01} = T_{10} = -\varepsilon_\textrm{b} \dot{a}ax, \\
   T_{11} &= \frac{\varepsilon_\textrm{b}\dot{a}^2 a^2 x^2
       + P_\textrm{b} a^2}{1-\dot{a}^2 x^2}, \mskip+12mu
      T_{22} = P_\textrm{b} a^2 x^2, \\
   T_{33} &= P_\textrm{b} a^2 x^2 \sin^2\theta.
\end{split}
\label{eq:tik_flat}
\end{equation}

From (\ref{eq:metric_flat}), (\ref{eq:raibk}) and (\ref{eq:tik_flat}),
the Einstein's field equations (\ref{eq:einstein}) reduce to
\begin{equation}
   P_\textrm{b}=-\varepsilon_\textrm{b}={{\Lambda}\over{8\pi G}}.
\label{eq:flat_evol}
\end{equation}
The $\varepsilon_\textrm{b}$ and $P_\textrm{b}$ should not be negative because
they are energy density and pressure of ordinary matter and radiation,
suggesting that
\begin{equation}
   \varepsilon_\textrm{b}=P_\textrm{b}=0 \mskip+12mu \textrm{and}
      \mskip+12mu \Lambda = 0. 
\end{equation}
Even if non-zero energies of matter and radiation with an equation
of state (\ref{eq:flat_evol}) can exist, the total energy density and
pressure should vanish:
\begin{equation}
   \varepsilon_\textrm{b}+\varepsilon_\Lambda = 0 \mskip+12mu
       \textrm{and} \mskip+12mu P_\textrm{b}+P_\Lambda = 0.
\end{equation}
Therefore, in the domain of classical physics the flat universe is empty,
which is consistent with the Einstein's claim that the infinite universe
has vanishing mean density \cite{ein22}.

Eq. (\ref{eq:flat_evol}) does not give any information about
the cosmic expansion history. 
Actually, the spatial homogeneity and isotropy condition
constrains the flat universe to be stationary.
In (\ref{eq:flat_tau}), the proper time interval of an arbitrary observer
depends on the choice of the comoving coordinate system.
The proper time of an observer moving faster at farther distance
from the origin goes slower.
Only at $x=0$ or if $\dot{a}=0$, it becomes $d\tau=d t\sqrt{1-a^2 v^2}$,
the same form of the proper time as in the special relativity,
irrespective of the choice of the comoving coordinate system.
Since the dependence of the proper time interval on the choice of reference
frame is contradictory to the spatial homogeneity and isotropy condition,
the flat universe should be stationary ($\dot{a}=0$).
In conclusion, the flat universe is empty and stationary,
and therefore is equivalent to the Minkowski space-time.

\subsection{Closed universe}
\label{sec:closed}

The homogeneous and isotropic closed space is usually described by
the spatially finite hypersphere with uniform and positive curvature.
By extending the example in Fig. \ref{fig:line}(b),
let us consider our universe as an expanding 3-sphere of curvature radius
$a(t)$, embedded in a 4-dimensional Euclidean space where each point
is labelled by the Cartesian coordinates $(x,y,z,w)$ and the world time $t$. 
The equation of the 3-sphere in $x$-$y$-$z$-$w$ coordinate system is 
\begin{equation}
   x^2 + y^2 + z^2 + w^2 = r^2 + w^2 = a^2,
\label{eq:sphere}
\end{equation}
where $r$ is the radial distance in $x$-$y$-$z$ coordinate system. 
The coordinates $x$, $y$, $z$ have transformation relations
with the spherical coordinates $r$, $\theta$, $\phi$ as 
$x=r\sin\theta\cos\phi$, $y=r\sin\theta\sin\phi$, and $z=r\cos\theta$.

The line element is defined as the space-time distance between
two infinitesimally separated events located at
$(t,x,y,z,w)$ and $(t+d t,x+d x,y+d y,z+d z,w+d w)$:
\begin{equation}
   d s^2 = d t^2 -d x^2 -d y^2 -d z^2 -d w^2 = d t^2 -d l^2,
\label{line:closed}
\end{equation}
where $d l^2$ is a spatial separation as measured in the world reference
frame. For two distinct events on the expanding 3-sphere,
the spatial separation is written as
\footnote{The 3-dimensional version of (\ref{eq:spsep_closed}), 
$dl^2=da^2 + a^2 (d\theta^2 + \sin^2 \theta d\phi^2)$,
is mathematically equivalent to Ref. \cite{dubrovin84} $\S9$(2)
that is a separation between two neighboring points in the 3-dimensional
Euclidean space, expressed in terms of spherical coordinates $(a,\theta,\phi)$.}
\begin{equation}
\begin{split}
   d l^2 &= dx^2 + dy^2 + dz^2 + dw^2 \\
         &= d r^2 + r^2 (d\theta^2 + \sin^2\theta d\phi^2)
            +{ {(a d a-r d r)^2} \over {a^2-r^2} } \\
         &= da^2 + a^2 \left[ d\chi^2 
            + \sin^2\chi (d\theta^2 + \sin^2\theta d\phi^2)\right],
\end{split}
\label{eq:spsep_closed}
\end{equation}
where $w$ has been removed by (\ref{eq:sphere}) and its differential
\begin{equation}
   rdr + wdw = ada = a\dot{a} dt,
\label{eq:rwa}
\end{equation}
and $r$ has been replaced with the comoving coordinate $\chi$
by a parametrization
\begin{equation}
   r = a \sin\chi \mskip+24mu (0\le\chi\le\pi),
\label{eq:r_chi}
\end{equation}
and its differential
\begin{equation}
   dr=\dot{a}\sin\chi dt + a \cos\chi d\chi.
\end{equation}
Therefore, the general form of line element for the non-stationary
closed universe is
\begin{equation}
\begin{split}
 ds^2  &= \left[1-\dot{a}^2(t)\right] dt^2  \\
       &\mskip+24mu -a^2(t)\left[ d\chi^2 + \sin^2\chi(d\theta^2 
                + \sin^2\theta d\phi^2) \right].
\end{split}
\label{eq:line_closed2}
\end{equation}
The metric tensor in the coordinate system $x^i=(t,\chi,\theta,\phi)$ is
\begin{equation}
   g_{ik} = \textrm{diag}\left[1-\dot{a}^2,-a^2,-a^2\sin^2\chi,
                          -a^2\sin^2\chi\sin^2\theta \right].
\label{eq:metric_closed}
\end{equation}
 
We calculate the Christoffel symbols and the Ricci tensor
from the metric tensor. The non-zero Christoffel symbols are
\begin{equation}
\begin{split}
   \Gamma^{0}_{00} &= {{-\dot{a}\ddot{a}}\over{1-\dot{a}^2}}, \mskip+12mu
      \Gamma^{0}_{11} = {{a\dot{a}}\over{1-\dot{a}^2}}, \mskip+12mu
      \Gamma^{0}_{22} = {{a\dot{a}}\over{1-\dot{a}^2}}\sin^2\chi, \\
   \Gamma^{0}_{33} &= {{a\dot{a}}\over{1-\dot{a}^2}}\sin^2\chi\sin^2\theta,
      \mskip+12mu \Gamma^{1}_{01} = \Gamma^{2}_{02} = \Gamma^{3}_{03}
      = {{\dot{a}}\over{a}}, \\
   \Gamma^{1}_{22} &= -\sin\chi\cos\chi, \mskip+12mu
      \Gamma^{1}_{33} = -\sin\chi\cos\chi\sin^2\theta, \\
   \Gamma^{2}_{12} &= \Gamma^{3}_{13}=\cot\chi, \mskip+12mu
      \Gamma^{2}_{33}= -\sin\theta\cos\theta, \mskip+12mu
      \Gamma^{3}_{23}= \cot\theta.
\end{split}
\end{equation}
The non-zero components of the Ricci tensor are
\begin{equation}
\begin{split}
   R_{00} &= -3\left( {{\ddot{a}}\over{a}}\right) {{1}\over{1-\dot{a}^2}}, \\
   R_{11} &= {{a\ddot{a}}\over{(1-\dot{a}^2})^2}+{{2}\over{1-\dot{a}^2}}, \\
   R_{22} &= R_{11}\sin^2\chi, \mskip+12mu R_{33}=R_{11}\sin^2\chi\sin^2\theta, 
\end{split}
\label{ricci:closed}
\end{equation}
and the Ricci scalar is
\begin{equation}
   R=-6\left( {{\ddot{a}}\over{a}}\right) {{1}\over{(1-\dot{a}^2)^2}}
     -6{{1}\over{a^2(1-\dot{a}^2)}}.
\label{ricci_scalar:closed}
\end{equation}

From (\ref{eq:line_closed2}), we obtain a proper time interval
as measured by an observer in arbitrary motion as 
\begin{equation}
   d\tau = d t\left(1-\dot{a}^2-a^2 v^2 \right)^{1/2},
\label{eq:dtau}
\end{equation}
and thus express the 4-velocity of the observer as
\begin{equation}
   u^{i}= \frac{d x^{i}}{d\tau} = \left(\gamma,\gamma \mathbf{v}\right),
\end{equation}
where 
\begin{equation}
   \gamma = (1-\dot{a}^2-a^2 v^2)^{-1/2}
\end{equation}
is a contraction factor. Note that the contraction factor depends on
the expansion speed of the universe as well as the peculiar motion
of the observer. 
For a comoving observer, the energy-momentum tensor for perfect fluid
is obtained by inserting the 4-velocity
$u^{i}=(1/\sqrt{1-\dot{a}^2},0,0,0)$ into (\ref{eq:emtensor}):
\begin{equation}
   T_{ik} = \textrm{diag}
            [\varepsilon_\textrm{b} (1-\dot{a}^2),P_\textrm{b} a^2,
            P_\textrm{b} a^2\sin^2\chi, 
            P_\textrm{b} a^2\sin^2\chi\sin^2\theta],
\label{em:closed}
\end{equation}
where $\varepsilon_\textrm{b}$ and $P_\textrm{b}$ are energy density and
pressure as defined in the world reference frame.
Inserting (\ref{eq:metric_closed}), (\ref{ricci:closed}),
(\ref{ricci_scalar:closed}) and (\ref{em:closed}) into (\ref{eq:einstein}),
we get evolution equations for homogeneous and isotropic
closed universe. They are concisely written as
\begin{equation}
   {{1}\over{(1-\dot{a}^2)^2}} {{\ddot{a}}\over{a}}
      = - {{4\pi G}\over{3}} (\varepsilon_\textrm{b} + 3P_\textrm{b})
        + {\Lambda \over 3}
\label{eq:evol_closed1}
\end{equation}
and
\begin{equation}
   {{1}\over{a^2(1-\dot{a}^2)}}
      = {{8\pi G}\over{3}} \varepsilon_\textrm{b} + {\Lambda \over 3}.
\label{eq:evol_closed2}
\end{equation}

Combining (\ref{eq:evol_closed1}) and (\ref{eq:evol_closed2}) gives
a continuity equation for energy density and pressure
\begin{equation}
   -3\left({\dot{a}}\over{a}\right)(\varepsilon+P)=\dot{\varepsilon},
\label{eq:conti}
\end{equation}
where $\varepsilon$ and $P$ are total energy density and pressure
of radiation (R), matter (M), and the cosmological constant ($\Lambda$): 
$\varepsilon = \sum_I \varepsilon_I$ and $P=\sum_I P_I$
($I=\textrm{R},\textrm{M},\Lambda$).
Note that $\varepsilon_\textrm{b} = \varepsilon_\textrm{R}
+ \varepsilon_\textrm{M}$.
The continuity equation is equivalent to $T^i_{~0;i}=0$, with a semicolon
denoting a covariant derivative.
It is worth noting that the time-time component of the metric tensor
($g_{00}=1-\dot{a}^2$) is positive according to the sign convention adopted.
The positiveness of the left-hand side of (\ref{eq:evol_closed2})
suggests that the total energy density should be positive
in the closed universe ($\varepsilon > 0$).

Introducing an equation of state $P_I=w_I \varepsilon_I$ for each species
$I$, we obtain a solution to (\ref{eq:conti}) as
$\varepsilon_I \propto a^{-3(1+w_I)}$. 
Thus the total energy density is written as
$\varepsilon = \sum_I \varepsilon_{I0} (a/a_0)^{-3(1+w_I)}$.
The subscript $0$ denotes the present epoch $t_0$.
Hereafter we call universes dominated by radiation, matter,
and the cosmological constant as radiation-universe (R-u),
matter-universe (M-u), and $\Lambda$-universe ($\Lambda$-u), respectively.
The energy density evolves as $\varepsilon_\textrm{R} \propto a^{-4}$ in the
radiation-universe ($w_\textrm{R} = \frac{1}{3}$),
$\varepsilon_\textrm{M} \propto a^{-3}$ in the matter-universe
($w_\textrm{M} = 0$),
and $\varepsilon_\Lambda = \textrm{const.}$ in the $\Lambda$-universe
($w_\Lambda = -1$). 
For the cosmological constant, $\varepsilon_\Lambda=\varepsilon_{\Lambda 0}$.

Let us define a dimensionless function of redshift $z \equiv a_0/a(t) -1$,
\begin{equation}
\begin{split}
   A^2(z) &\equiv \frac{a_0^2}{a^2 (1-\dot{a}^2)}  \\
      &= {{8\pi G a_0^2}\over{3}} \left[ \varepsilon_{\textrm{R}0} (1+z)^4  
      + \varepsilon_{\textrm{M}0}(1+z)^3 + \varepsilon_\Lambda \right] \\
      &= a_0^2 \left[ \frac{(1+z)^4}{a_{\textrm{R}0}^2}
                        +\frac{(1+z)^3}{a_{\textrm{M}0}^2}
                        +\frac{1}{a_{\Lambda}^2}\right],
\end{split}
\label{eq:az2}
\end{equation}
where $a_{I0}=(3/8\pi G\varepsilon_{I0})^{1/2}$.
Throughout this paper, a function of time $t$ will be expressed
in terms of redshift $z$ interchangeably. 
The radius parameter $a_{\textrm{R}0}$ ($a_{\textrm{M}0}$) can be interpreted
as a free-fall time or radius for the gravitational collapse of a stationary 
radiation- (matter-) universe with the present radiation (matter)
energy density, and $a_\Lambda = a_{\Lambda 0}$ as the minimum radius
of $\Lambda$-universe. By introducing another dimensionless quantity
\begin{equation}
   D_I \equiv {{8\pi G\varepsilon_I}\over 3} a^2 (1-\dot{a}^2)
               = \left(a_0\over a_{I0}\right)^2 
                 {{(1+z)^{3(1+w_I)}}\over{A^2(z)}},
\label{eq:di}
\end{equation} 
we can rewrite (\ref{eq:evol_closed2}) as 
\begin{equation}
   D_\textrm{R} + D_\textrm{M} + D_\Lambda = 1,
\label{eq:ddd}
\end{equation}
which holds during the whole history of the universe.
The $D_I$ can be interpreted as the fraction of energy of species $I$
(see Sec. \ref{sec:energy}).

\subsection{Open universe}
\label{sec:open}

We now consider the geometry of homogeneous and isotropic expanding
3-dimensional space with uniform and negative curvature.
Such a negatively curved space cannot be embedded in a 4-dimensional
Euclidean space. At an instant of time, it is a pseudosphere with imaginary
radius $ia$ ($\S111$ of Ref. \cite{lan75}).
We replace $a^2$ with $-a^2$ in (\ref{eq:sphere}) to obtain an expression
analogous to (\ref{eq:spsep_closed}) for a spatial separation between
two distinct events on the expanding 3-pseudosphere
\footnote{The 3-dimensional hypersurface with uniform and negative curvature
can also be represented as a pseudosphere of radius $a$,
satisfying the equation $w^2 - x^2 - y^2 - z^2 = a^2$,
embedded in a 4-dimensional pseudo-Euclidean space $\mathbb{R}_1^4$.
Here, the spatial separation between two neighboring points
on the non-stationary pseudosphere has the same form as (\ref{eq:spsep_open})
aside from the sign, $dl^2 = dw^2 - dx^2 - dy^2 - dz^2 = da^2 - a^2
[d\chi^2 + \sinh^2\chi (d\theta^2 + \sin^2\theta d\phi^2)]$.
See Ref. \cite{dubrovin84} $\S3.2$(12) for the 3-dimensional case
in $\mathbb{R}_1^3$.},
\begin{equation}
\begin{split}
   d l^2 &= d x^2 + d y^2 + d z^2 + d w^2 \\
        &= d r^2 + r^2 (d\theta^2 + \sin^2 \theta d\phi^2 )
           - \frac{(a d a+r d r)^2}{a^2 + r^2} \\
        &= -da^2 + a^2 \left[ d\chi^2 + \sinh^2\chi
           (d\theta^2 + \sin^2\theta d\phi^2)\right],
\end{split}
\label{eq:spsep_open}
\end{equation}
where the radial distance $r=(x^2 + y^2 + z^2)^{1/2}$ in $x$-$y$-$z$
coordinate system has been parametrized with the comoving coordinate $\chi$ by
\begin{equation}
   r = a \sinh\chi \mskip+24mu (\chi \ge 0).
\end{equation}
Therefore, the line element for the non-stationary open universe is 
\begin{equation}
\begin{split}
 d s^2 &= \left[1+\dot{a}^2(t)\right]d t^2  \\
       &\mskip+24mu -a^2(t)\left[ d\chi^2 + \sinh^2\chi(d\theta^2
            + \sin^2\theta d\phi^2)\right].
\end{split}
\label{eq:line_open}
\end{equation}

The non-zero components of the Ricci tensor calculated from the metric
(\ref{eq:line_open}) are
\begin{equation}
\begin{split}
   R_{00} &= -3\left( {{\ddot{a}}\over{a}}\right) {{1}\over{1+\dot{a}^2}}, \\
   R_{11} &= {{a\ddot{a}}\over{(1+\dot{a}^2})^2}-{{2}\over{1+\dot{a}^2}}, \\
   R_{22} &= R_{11}\sinh^2\chi, \mskip+12mu
      R_{33}=R_{11}\sinh^2\chi\sin^2\theta, 
\end{split}
\label{ricci:open}
\end{equation}
and the Ricci scalar is
\begin{equation}
   R=-6\left( {{\ddot{a}}\over{a}}\right) {{1}\over{(1+\dot{a}^2)^2}}
     +6{{1}\over{a^2(1+\dot{a}^2)}}.
\label{ricci_scalar:open}
\end{equation}
For a comoving observer with 4-velocity $u^{i}=(1/\sqrt{1+\dot{a}^2},0,0,0)$,
the energy-momentum tensor for perfect fluid becomes
\begin{equation}
\begin{split}
   T_{ik} &= \textrm{diag}
            [\varepsilon_\textrm{b} (1+\dot{a}^2),P_\textrm{b} a^2, \\
          & \mskip+60mu P_\textrm{b} a^2\sinh^2\chi, 
            P_\textrm{b} a^2\sinh^2\chi\sin^2\theta].
\end{split}
\label{em:open}
\end{equation}

The resulting evolution equations for the open universe are obtained
in the same way as those for the closed universe are obtained. They are
\begin{equation}
   {{1}\over{(1+\dot{a}^2)^2}} {{\ddot{a}}\over{a}}
      = - {{4\pi G}\over{3}} (\varepsilon_\textrm{b}+3P_\textrm{b})
        + {\Lambda \over 3}
\label{eq:open1}
\end{equation}
and
\begin{equation}
   {{1}\over{a^2(1+\dot{a}^2)}}
      = -{{8\pi G}\over{3}} \varepsilon_\textrm{b} - {\Lambda \over 3}.
\label{eq:open2}
\end{equation}
Combining (\ref{eq:open1}) and (\ref{eq:open2}) also gives
the same continuity equation as (\ref{eq:conti}).
Since the time-time component of the metric tensor ($g_{00}=1+\dot{a}^2$)
is always positive, equation (\ref{eq:open2}) suggests that the total energy
density should be negative in the open universe ($\varepsilon < 0$).

\subsection{Our universe is spatially closed}
\label{sec:ousc}

In Secs. \ref{sec:flat}--\ref{sec:open}, we have demonstrated that
flat universe is equivalent to the Minkowski space-time, which is empty
and stationary, and that closed and open universes have positive and negative
energy densities, respectively. In other words, the curvature of the universe
is determined by the sign of mean energy density, not by the ratio of
the energy density to the critical density as in the FRW world. 
The open universe is unrealistic because the mean density of the universe
is known to be positive from astronomical observations. 
Therefore, we conclude that our universe is spatially closed and finite.
The spatial closure of the universe has been deduced from the purely
theoretical point of view.
Our conclusion verifies the Einstein's claim for the finiteness of the universe.

\subsection{The Friedmann equations}
\label{sec:rel_friedmann}

In the non-flat universes, the proper time interval of a comoving observer
is related to the world time by
\begin{equation}
   d\tau = \left[ 1 - k \dot{a}^2 (t) \right]^{1/2} d t,
\label{eq:tau}
\end{equation}
where $k=+1$ for closed and $-1$ for open universes.
The proper time goes slower (faster) than the world time in the closed (open)
universe, with non-uniform lapse.
Only in the flat universe, the proper time elapses uniformly
($d\tau=d t$).

From (\ref{eq:tau}), we obtain relations between world- and
proper-time derivatives of the curvature radius:
\begin{equation}
   \frac{1}{a^2}\left(\frac{d a}{d\tau}\right)^2 + \frac{k}{a^2}
      = \frac{k}{a^2 (1-k \dot{a}^2)}
\label{eq:frw_new1}
\end{equation}
and
\begin{equation}
   \frac{1}{a}\left(\frac{d^2 a}{d\tau^2}\right)
      = \frac{1}{(1-k \dot{a}^2)^2}\frac{\ddot{a}}{a},
\label{eq:frw_new2}
\end{equation}
where the proper time $\tau$ acts as the time of non-flat FRW world models.
As will be shown in Sec. \ref{sec:energy} [Eq. (\ref{eq:medensity})],
energy density and pressure defined in the world reference frame
are equivalent to those measured by the comoving observer.
Therefore, Friedmann equations (\ref{eq:frw1}) and (\ref{eq:frw2})
for non-flat universes can be derived from the new cosmic evolution
equations [Eqs. (\ref{eq:evol_closed1}) and (\ref{eq:evol_closed2})
for closed and (\ref{eq:open1}) and (\ref{eq:open2}) for open universes]
by the time-parametrization (\ref{eq:tau}).
This, along with the local nature of the FRW metric as discussed in Sec.
\ref{sec:metric}, implies that the Friedmann equations describe
the evolution of the local universe around a comoving observer.

\begin{table}
\caption{\label{tab:E_sign}Possible ranges of total energy density in
         the FRW and new world models.}
\begin{ruledtabular}
\begin{tabular}{ccc}
 Curvature Type  & FRW World\footnotemark[1] & New World  \\
\hline\\[-3mm]
 closed & $\varepsilon \ge 3/{8\pi G a^2}$ & $\varepsilon >0$ \\
 flat   & $\varepsilon \ge 0$ & $\varepsilon = 0$ \\
 open   & $ -3/{8\pi G a^2} \le \varepsilon < 0$ & $\varepsilon <0$ \\
\end{tabular}
\end{ruledtabular}
\footnotetext[1]{Equality signs for the FRW world models
                 correspond to cases of stationary universes
                 ($\dot{a}=0$, $\ddot{a}\ne 0$).}
\end{table}

Table \ref{tab:E_sign} lists possible ranges of total energy density
in the FRW and new world models.
In the new world models, the energy density is strictly positive, zero,
and negative for closed, flat, and open universes, respectively.
On the other hand, the FRW world models have rather complicated ranges
of energy density.
From (\ref{eq:frw_new1}), one finds that $(d a/d\tau)^2 \ge 0$
for closed, and $0 \le (d a/d\tau)^2 < 1$ for open universes.
Thus, Eq. (\ref{eq:frw2}) plus constraints on $(d a/d\tau)^2$
suggests that the closed FRW world model have positive energy density
larger than or equal to $3/8\pi G a^2$,
and that the open model accommodate only negative energy density
not smaller than $-3/8\pi G a^2$.
Note that the positive energy density appears to be allowable
in the open FRW world model if the constraint on $(d a/d\tau)^2$ is not imposed.

The Friedmann equations for the flat universe ($k=0$) cannot be
derived with any world-proper time relation, but can be obtained
by neglecting the curvature term $-k/a^2$ in (\ref{eq:frw2})
for the closed model, resulting in $\varepsilon \ge 0$.
The flat FRW world model is valid only within regions where the effect
of curvature is negligible or the distance scale of interest is far smaller
than the curvature radius.
In other words, the flat FRW world is an approximation of
the closed universe with large curvature radius.
Its hypersurface is a tangent space of a comoving observer on the $3$-sphere.

All the differences between FRW and new world models come from
a difference between reference frames adopted.
In later sections of this paper, to describe physical phenomena
in the expanding closed universe, we use both the world reference frame
and the comoving observer's frame.
The former has a global time $t$ with uniform lapse, while the latter
has a local time $\tau$ whose lapse depends on the cosmic expansion speed.
An event on the expanding 3-sphere may be labelled by coordinates
$(t,x,y,z,w)$ or $(t,a,\chi,\theta,\phi)$.
By omitting the curvature radius $a$ that is a function of $t$, we regard
the coordinate system $x^i = (t,\chi,\theta,\phi)$ as equivalent
to the world reference frame.
The comoving observer's frame is a system of physical space and
proper time coordinates adopted by an observer like us whose comoving
coordinate $x^\alpha = (\chi,\theta,\phi)$ is fixed during the expansion
of the universe. Actually, the comoving observer's frame is equivalent
to the locally inertial frame.

\section{Physical and astronomical aspects of the expanding closed universe}
\label{sec:some}

In this section, we investigate interesting properties of the expanding
closed universe, such as time-varying light speed, cosmic expansion history,
energy-momentum relation of particles, redshift, and cosmic distance
and time scales.
All the quantities, not otherwise specified, are defined and expressed
in the world reference frame, and secondarily in the comoving observer's
frame. In the latter frame, all the physical quantities and their evolution
are the same as those in the closed FRW world.

\subsection{Time-varying light speed and cosmic expansion speed} 
\label{sec:speed}

In the special theory of relativity, the speed of light is constant and
equal to the limiting speed ($c=1$), which applies to the Minkowski space-time,
or equivalently to the flat universe.
In the expanding closed universe, however, the light speed is less than
or equal to the limiting speed. 
From the photon's geodesic equation ($d s=0$),
one can express the speed of light as
\begin{equation}
    \eta(t) \equiv (1-\dot{a}^2)^{1/2}={(1+z)\over {A(z)}}. 
\label{eq:eta_closed}
\end{equation}
It should be noted that the light speed varies with time, depending
on the cosmic expansion speed $\dot{a}$ and satisfying $\eta^2+\dot{a}^2=1$.
Both speeds cannot exceed the limiting speed ($0 \le \eta \le 1$
and $-1 \le \dot{a} \le 1$). 
The cosmic expansion speed goes to unity as the redshift goes to infinity 
because $A(z)$ evolves as $(1+z)^2$ in the radiation-dominated era
[Eq. (\ref{eq:az2})]. Therefore, we expect $\dot{a}=1$ and $\eta=0$
at the beginning of the universe.

On the other hand, the comoving observer always measures the speed of light
as unity because the observer's proper time interval varies in the same way
as the world-frame light speed does 
[Eqs. (\ref{eq:tau}) and (\ref{eq:eta_closed})]. 
Besides, the cosmic expansion speed in the comoving observer's frame
has no limit ($d a/d\tau = \dot{a}/\sqrt{1-\dot{a}^2} < \infty$).

In the open universe, the speed of light is $\eta=(1+\dot{a}^2)^{1/2} \ge 1$:
the light propagates faster than the limiting speed. 
There is no upper limit on the cosmic expansion and the light speeds
in the world reference frame. However, the comoving observer perceives that
the speed of light is always unity and that the cosmic expansion speed
is bounded to unity ($d a/d\tau=\dot{a}/\sqrt{1+\dot{a}^2} < 1$),
as discussed in Sec. \ref{sec:rel_friedmann}.

\subsection{Cosmic evolution history}
\label{sec:evhist}

The evolution of homogeneous and isotropic universe is described
by the evolution of physical quantities during the history of the universe.
The most important quantity is the curvature radius $a$.
Although it is not easy to get the general solution to (\ref{eq:evol_closed2}),
there exist analytic solutions for special cases of the universe dominated
by energy of the single species.
For radiation-universe ($\varepsilon_\textrm{M} = 0$, $\varepsilon_\Lambda = 0$),
the curvature radius is given by 
\begin{equation}
   a(t)=b_\textrm{R} \sin(t/b_\textrm{R}) \mskip+24mu (0 \le t \le \pi b_\textrm{R}),
\label{eq:rde}
\end{equation}
where $b_\textrm{R}=\left( 8\pi G \varepsilon_{\textrm{R}0} a_0^4/3\right)^{1/2}
=(a_0 / a_{\textrm{R}0})a_0$ is the maximum curvature radius of
the radiation-universe.
At $t=0$, the universe expands with the maximum speed and
zero acceleration ($\dot{a}=1$, $\ddot{a}=0$).
The positive acceleration is not allowable in the radiation-universe.

For matter-universe ($\varepsilon_\textrm{R}=0$, $\varepsilon_\Lambda = 0$),
the solution for the curvature radius is 
\begin{equation}
  a(t)=b_\textrm{M} - {1 \over {4 b_\textrm{M}}} (t-2 b_\textrm{M})^2
     \mskip+24mu (0 \le t \le 4 b_\textrm{M}),
\label{eq:mde}
\end{equation}
where $b_\textrm{M}=8\pi G\varepsilon_{\textrm{M}0} a_0^3/3
=(a_0/a_{\textrm{M}0})^2 a_0$ is the maximum curvature radius of
the matter-universe. 
The initial condition $a(0)=0$ has been assumed.
The cosmic expansion acceleration is negatively constant in
the matter-universe ($\ddot{a}=-1/2b_\textrm{M}$).
Due to the negative acceleration, the expanding universe containing
only radiation and matter is bound to contract into the single point.

If the universe does not contain the ordinary matter and radiation
but is dominated by the cosmological constant or dark energy
($\Lambda$-universe), Eq. (\ref{eq:evol_closed2}) becomes
\begin{equation}
 {{d a}\over{d t}}=\pm \sqrt{1-{{3}\over{\Lambda a^2}}} 
    = \pm \sqrt{1 - \left({a_{\Lambda}}\over{a}\right)^2}, 
\label{eq:adot_dde}
\end{equation}
where $a_{\Lambda} \equiv (3/\Lambda)^{1/2}
= (3/8\pi G \varepsilon_{\Lambda})^{1/2}$ is the (minimum) radius
of the $\Lambda$-universe at initial time $t_i$.
For the expanding $\Lambda$-universe, we get
\begin{equation}
   a(t) = \left[ a_{\Lambda}^2 + (t-t_i)^2 \right]^{1/2}\mskip+24mu (t \ge t_i).
\label{eq:dde}
\end{equation}
The expansion speed and acceleration of the $\Lambda$-universe are
$\dot{a}(t) = (t-t_i)/[ a_{\Lambda}^2 + (t-t_i)^2 ]^{1/2}$
and $\ddot{a}(t)=a_{\Lambda}^2 / [ a_{\Lambda}^2 + (t-t_i)^2 ]^{3/2}$,
which go over asymptotically into unity and zero, respectively,
as $t$ goes to infinity.
Starting with $a(t_i)=a_{\Lambda}$, $\dot{a}(t_i)=0$, and 
$\ddot{a}(t_i)=a_\Lambda^{-1}$, the $\Lambda$-universe expands eternally.

It is interesting to consider a universe dominated by energy of
a hypothetical species with an equation of state $w_\textrm{H}=-\frac{1}{3}$
\cite{kolb89}. 
This universe (H-universe) has a simple expansion history
\begin{equation}
 a(t)=t [1-(a_{\textrm{H}0}/a_0)^2]^{1/2} \mskip+24mu (t \ge 0),
\label{eq:hde}
\end{equation}
where $a_{\textrm{H}0} = (3/8\pi G \varepsilon_{\textrm{H}0})^{1/2}$. 
The important property of the H-universe is that
$a \propto t$ and $\ddot{a} = 0$.
The universe expands with the constant speed. 
The energy density of the hypothetical species evolves as
$\varepsilon_\textrm{H} \propto a^{-2}$.
For an extreme case of $a_{\textrm{H}0}=0$ ($\varepsilon_{\textrm{H}0} = \infty$),
the H-universe expands with the limiting speed ($a=t$).

\begin{table*}
\caption{Cosmological parameters of the two closed world models.}
\begin{tabular}{lccc}
\hline\hline
Parameters  & Symbols & Model I\footnotemark[1] & Model II\footnotemark[2]  \\
\hline 
\multicolumn{4}{c}{Input} \\
\hline \\[-3mm]
 CMB temperature     & $T_\textrm{cmb}$ & $2.725$ K  & $2.725$ K  \\
 Matter density      & $\Omega_\textrm{M}$ & $0.415$  & $0.315$  \\
 Dark energy density & $\Omega_{\Lambda}$ & $0.630$  & $0.696$  \\
 Hubble constant     & $h$ & $0.55$ & $0.71$  \\[1mm]
\hline
\multicolumn{4}{c}{Derived} \\
\hline \\[-3mm]
 Curvature radius  & $a_0$ & 
                       $25670$ Mpc & $40170$ Mpc \\
 Speed of light    & $\eta_0 /c$    & 
                       $0.208$  & $0.105$  \\
 Expansion speed   & $\dot{a}_0 /c$ & $0.978$ & $0.995$ \\
 Expansion acceleration   & $\ddot{a}_0$ &
                          $6.79\times 10^{-7}$ Mpc$^{-1}$  &
                          $1.45\times 10^{-7}$ Mpc$^{-1}$  \\
 Cosmic age (in world time)   & $t_0$  & $85.3$ Gyr & $132$ Gyr \\
 Cosmic age (in proper time)  & $\tau_0$   & $15.8$ Gyr & $13.1$ Gyr \\[2mm]
 Radiation free-fall radius   & $a_{\textrm{R}0}$ & $603.1$ Gpc & $603.1$ Gpc \\
 Matter free-fall radius      & $a_{\textrm{M}0}$ & $8461$ Mpc & $7523$ Mpc \\
 Minimum radius of $\Lambda$-u  & $a_{\Lambda}$ & $6867$ Mpc & $5061$ Mpc \\
 Maximum radius of R-u          & $b_\textrm{R}$ & $1093$ Mpc & $2675$ Mpc \\
 Maximum radius of M-u          & $b_\textrm{M}$ & $236.3$ Gpc & $1145$ Gpc \\[2mm]
 3-volume of the universe & $V_0$ &
                        $6.94\times 10^{13}$ $\textrm{Mpc}^3$ &
                        $1.34\times 10^{14}$ $\textrm{Mpc}^3$ \\
 Total radiation energy & $E_{\textrm{R}0}$ & 
                        $8.50\times 10^{74}$ erg  &
                        $1.64\times 10^{75}$ erg \\
 Total matter energy    & $E_{\textrm{M}0}$ & 
                        $4.32\times 10^{78}$  erg &
                        $1.05\times 10^{79}$  erg\\
 Total dark energy      & $E_{\Lambda 0}$ & 
                        $6.56\times 10^{78}$ erg &
                        $2.33\times 10^{79}$ erg \\[2mm]
 Fraction of radiation energy        & $D_{\textrm{R}0}$ & 
                        $7.82\times 10^{-5}$ & $4.85\times 10^{-5}$ \\
 Fraction of matter energy & $D_{\textrm{M}0}$ & $0.397$ & $0.312$ \\
 Fraction of dark energy & $D_{\Lambda 0}$ & $0.603$ & $0.688$ \\[1mm]
\hline\hline
\end{tabular}
\vskip 0.3cm \hskip 0.25cm {\begin{minipage}{0.8\linewidth}
{\small \setlength{\baselineskip}{1pt}
\footnotetext[1]{
       Based on parameters of a non-flat FRW world model
       that best fits with the WMAP 3-year data only \cite{spergel07}.}
\footnotetext[2]{
       Based on parameters of a non-flat FRW world model
       that jointly fits with the CMB, SNIa, $\gamma$-ray bursts,
       the shape parameter $\Gamma=\Omega_\textrm{M} h$, Hubble constant,
       matter density ($\Omega_\textrm{M} h^2$), and
       big-bang nucleosynthesis data \cite{wright07}.
       The Hubble constant is an average of recent measurement values.}
}
\end{minipage}}
\label{tab:cospar}
\end{table*}

To reconstruct the evolution history of the closed universe, we have
adopted two world models that are consistent with the recent astronomical
observations. The model parameters, which are listed in Table \ref{tab:cospar},
are based on a non-flat $\Lambda\textrm{CDM}$ FRW world model
that best fits with the WMAP CMB data only
(Model I; $H_0=55$ km s$^{-1}$ Mpc$^{-1}$, $\Omega_\textrm{M} = 0.415$,
$\Omega_\Lambda = 0.630$; $\S3.3$ of \cite{spergel07})
and on another model that jointly fits with the CMB and other astronomical
data (Model II; $H_0=71$ km s$^{-1}$ Mpc$^{-1}$, $\Omega_\textrm{M} = 0.315$,
$\Omega_\Lambda = 0.696$; \cite{wright07}),
where $H_0=100h$ km s$^{-1}$ Mpc$^{-1}$ is the Hubble constant,
and $\Omega_I$ is the current density parameter of the FRW world model.
The Hubble constant of Model I is quite lower than the popular value of
Model II, but is allowable because the low Hubble constant has been
reported from observations of Cepheids plus SNIa
($H_0=62.3\pm 1.3\pm5.0$ km s$^{-1}$ Mpc$^{-1}$; \cite{sandage06})
and of eclipsing binaries ($H_0=61$ km s$^{-1}$ Mpc$^{-1}$;
\cite{bonanos06}).

Using the FRW model parameters as input, we calculate radius parameters
$a_{\textrm{R}0}$, $a_{\textrm{M}0}$, and $a_\Lambda$ from a formula
$a_{I0} = (3/8\pi G\varepsilon_{I0})^{1/2}=H_0^{-1}\Omega_I^{-1/2}$.
The radiation energy density has been calculated
from $\varepsilon_{\textrm{R}0} = \pi^2 k_\textrm{B}^4 T_\textrm{cmb}^4 /15\hbar^3$
(see Sec. \ref{sec:relgas}) with the CMB temperature
$T_\textrm{cmb}=2.725$ K \cite{mather99}.
For simplicity, the neutrino contribution to the radiation energy density
has been omitted.
The present curvature radius $a_0$ has been obtained from the relation
\begin{equation}
   H_0 = a_0^{-1} (d a/d\tau)_0
       = \left(a_{\textrm{R}0}^{-2}+a_{\textrm{M}0}^{-2}
        +a_\Lambda^{-2}-a_0^{-2}\right)^{1/2}.
\end{equation}
The basic parameters characterizing the closed world model are
the curvature radius of the universe ($a_0$) and the radius parameters
($a_{\textrm{R}0}$, $a_{\textrm{M}0}$, $a_\Lambda$) at the present time.

It is interesting to note that converting parameters of the flat FRW world
model into those of the new closed model always gives limiting values
of $a_0 = \infty$, $\dot{a}_0 = 1$, and $\ddot{a}_0 = 0$.
The flat FRW world is a limiting case of the closed universe with infinite
curvature radius.

\begin{figure}
\mbox{\epsfig{file=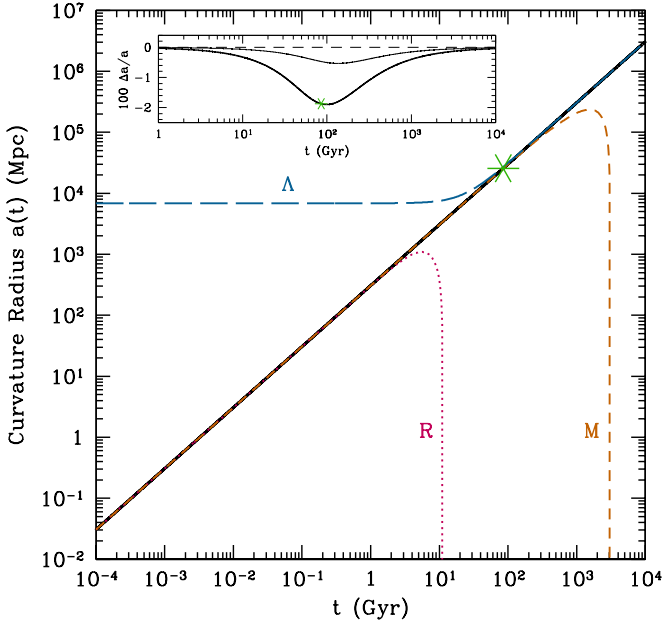,width=88mm,clip=}}
\caption{Evolution of curvature radius $a$ over the world time $t$
         (thick solid curve; Model I).
         For comparison, the curvature radius of radiation, matter, and
         $\Lambda$-universes are shown as dotted (R), dashed (M), and
         long-dashed ($\Lambda$) curves, respectively.
         The small panel shows the fractional difference of the curvature
         radius of our universe relative to $a=t$
         ($\Delta a/a\equiv[a-t]/t$ in unit of percent; 
         thick and thin solid curves for Model I and II, respectively).
         The stars denote quantities at the present time
         $t_0=85.3$ Gyr (Model I; see Sec. \ref{sec:cosmic_scales}).
         A rough estimate of redshift is given by $z\approx t_0/t-1$.}
\label{fig:aplot}
\end{figure}

The evolution histories of the two closed world models are summarized 
in Figs. \ref{fig:aplot}--\ref{fig:accplot} below, where we have
also plotted analytic solutions for radiation, matter, and $\Lambda$-universes
($t_i = 0$ is assumed for $\Lambda$-u).
All the numerical values given in the text are based on Model I.
The evolution of curvature radius of the closed world model
is shown in Fig. \ref{fig:aplot}, where we have performed a numerical
integration of (\ref{eq:evol_closed2}) to obtain $a(t)$.
Note that the solution of H-universe with infinite energy density
greatly approximates the evolution of curvature radius of our universe,
which differs from $a=t$ by maximally about $2$\% at $t\simeq 100$ Gyr,
as shown in the small panel.

\begin{figure}
\mbox{\epsfig{file=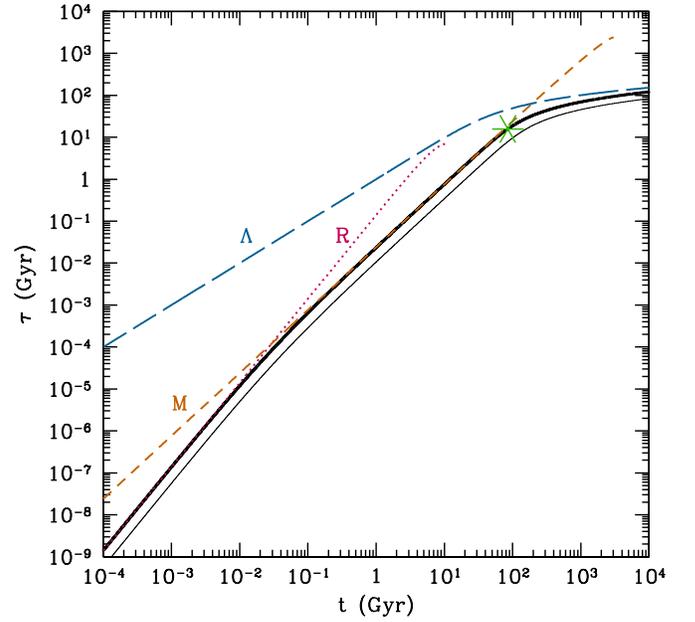,width=88mm,clip=}}
\caption{The proper time of a comoving observer $\tau$ versus
         the world time $t$ (thick and thin solid curves for Model I and II,
         respectively).
         The $\tau$-$t$ relations expected in radiation, matter,
         and $\Lambda$-universes are shown as dotted (R),
         dashed (M), and long-dashed ($\Lambda$) curves, respectively.
         The star indicates $t_0 = 85.3$ Gyr and $\tau_0=15.8$ Gyr.}
\label{fig:tplot}
\end{figure}

Fig. \ref{fig:tplot} shows the relation between the proper time of
a comoving observer and the world time, obtained by integrating (\ref{eq:tau}).
The total elapsed time until the present time, the age of the universe,
is denoted as a star at $t_0 = 85.3$ Gyr and $\tau_0 = 15.8$ Gyr
(see Sec. \ref{sec:cosmic_scales}).
The $\tau$-$t$ relations expected in radiation, matter, and
$\Lambda$-universes are written in analytic forms as
\begin{equation}
   \tau = b_\textrm{R} \left[ 1-\cos(t/b_\textrm{R}) \right] \mskip+24mu
                       (\textrm{R-u}),
\label{eq:tau_rde}
\end{equation}
\begin{equation}
\begin{split}
   \tau &= b_\textrm{M}
      \Bigg[\left({{t-2 b_\textrm{M}}\over{2 b_\textrm{M}}}\right) 
      \sqrt{1-\left({{t-2 b_\textrm{M}}\over{2 b_\textrm{M}}}\right)^2}  \\
    &\mskip+60mu + {\pi\over 2}
        + \arcsin\left({{t-2 b_\textrm{M}}\over{2 b_\textrm{M}}}\right)\Bigg]
        \mskip+24mu (\textrm{M-u}), 
\end{split}
\label{eq:tau_mde}
\end{equation}
and
\begin{equation}
   \tau = a_{\Lambda} \ln\left[\frac{t-t_i}{a_{\Lambda}} 
        + \sqrt{1+\left(\frac{t-t_i}{a_{\Lambda}}\right)^2}\right]
         \mskip+24mu (\Lambda\textrm{-u}),
\label{eq:tau_dde}
\end{equation}
respectively.
Inserting (\ref{eq:tau_rde}) into (\ref{eq:rde}) gives
$a(\tau)=[\tau (2 b_\textrm{R} -\tau)]^{1/2}$, which goes over into
$a \propto \tau^{1/2}$ if $\tau \ll 2b_\textrm{R}$, the behavior of
a scale factor in the radiation-dominated FRW universe.

\begin{figure}
\mbox{\epsfig{file=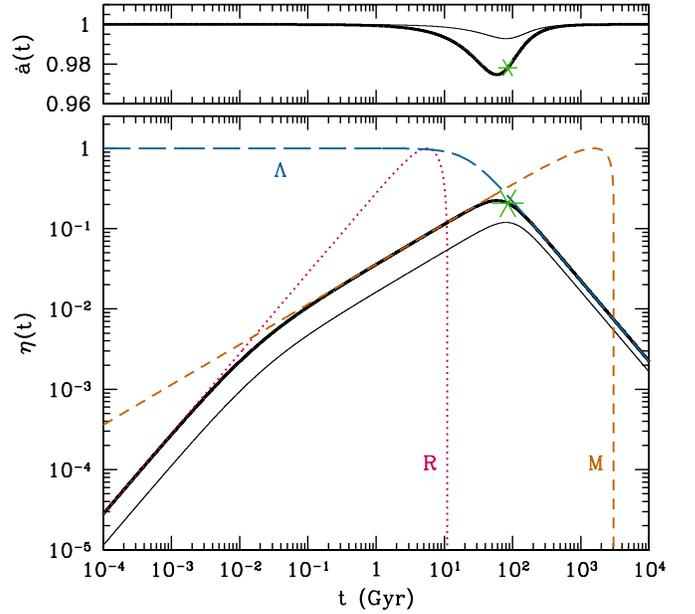,width=88mm,clip=}}
\caption{Variation of (top) cosmic expansion speed $\dot{a}$ and
         (bottom) speed of light $\eta$ over the world time $t$
         (in unit of the limiting speed of the special relativity;
         thick and thin solid curves for Model I and II, respectively).
         The relation $\eta^2 + \dot{a}^2 =1$ holds.
         The speed of light in radiation, matter, and $\Lambda$-universes
         are shown as dotted (R), dashed (M), and long-dashed ($\Lambda$)
         curves. The stars denote quantities at the present time.}
\label{fig:adotplot}
\end{figure}

Fig. \ref{fig:adotplot} shows the time-variation of cosmic expansion
speed (top) and speed of light (bottom).
At the present time, the universe is expanding faster than the light
by a factor of $4.7$ with $\dot{a}=0.978$ and $\eta=0.208$ (denoted as stars).
The speed of light in radiation, matter, and $\Lambda$-universes are written as
\begin{equation}
   \eta(t) = \sin (t/b_\textrm{R}) = a/b_\textrm{R} \mskip+24mu (\textrm{R-u}),
\label{eq:eta_rde}
\end{equation}
\begin{equation}
   \eta(t) = \sqrt{1-\left(\frac{t-2b_\textrm{M}}{2b_\textrm{M}}\right)^2}
           = (a/b_\textrm{M})^{1/2} \mskip+24mu (\textrm{M-u}),
\label{eq:eta_mde}
\end{equation}
and
\begin{equation}
   \eta(t) = \frac{a_\Lambda}{\left[a_\Lambda^2 + (t-t_i)^2 \right]^{1/2}}
           = a_{\Lambda}/a \mskip+24mu (\Lambda\textrm{-u}),
\label{eq:eta_dde}
\end{equation}
respectively. The behavior of time-varying light speed implies that
photons are frozen ($\eta=0$) when the universe expands with the maximum speed,
e.g., at the beginning or far in the future of the universe. 

\begin{figure}
\mbox{\epsfig{file=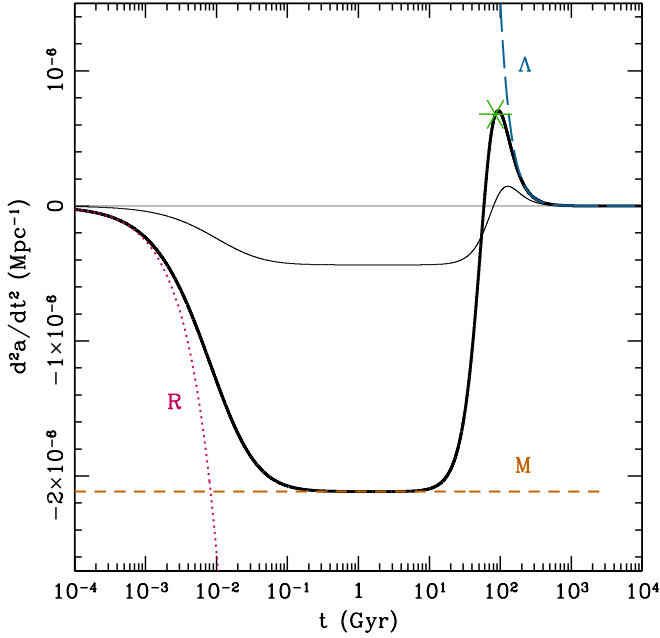,width=88mm,clip=}}
\caption{Evolution of cosmic expansion acceleration $\ddot{a}$ in unit of
         $\textrm{Mpc}^{-1}$ 
         (thick and thin solid curves for Model I and II, respectively).
         The acceleration in radiation, matter, and $\Lambda$-universes
         are shown as dotted (R), dashed (M), and long-dashed ($\Lambda$)
         curves, respectively.
         The star denotes the acceleration at the present time.}
\label{fig:accplot}
\end{figure}

Fig. \ref{fig:accplot} shows the history of cosmic expansion acceleration
calculated from
\begin{equation}
   \ddot{a} = - \frac{3(\varepsilon + 3P)}{16\pi G \varepsilon^2 a^3},
\end{equation}
which has been obtained by combining (\ref{eq:evol_closed1}) and
(\ref{eq:evol_closed2}).
In the radiation-dominated era, the expansion acceleration had gradually
decreased from zero, and became negatively constant
during the matter-dominated era ($t=0.1$--$10$ Gyr).
Afterward, the universe has been decelerated until $t=58.8$ Gyr
($\tau=10.0$ Gyr, $z=0.448$) from which it starts to accelerate
positively. The transition epoch corresponds to the point of maximum (minimum)
speed of light (expansion speed).
The universe arrives at the maximum acceleration at $t=93.5$ Gyr
($\tau=17.5$ Gyr, $z=-0.088$).
The present universe is on a stage before the maximum acceleration.
The future universe will expand eternally with asymptotic acceleration
of zero.

\subsection{Energy-momentum relation of particles}
\label{sec:energy_momentum}

Now we define energy and momentum of a free particle with rest mass $m$
in the closed universe.
Here, we mean the rest mass by the intrinsic mass of the particle 
that is independent of its peculiar motion and the dynamics of the universe.
Thus $m$ is the mass as measured in a locally inertial frame
comoving with the particle.
It is also equivalent to the rest mass in the stationary universe.
The action for the free material particle moving along a trajectory
with end points $\textrm{A}$ and $\textrm{B}$ has the form
($\S8$ of Ref. \cite{lan75})
\begin{equation}
   S = -m \int_\textrm{A}^\textrm{B} d s = \int_{t_\textrm{A}}^{t_\textrm{B}} L d t,
\end{equation}
where the Lagrangian
\begin{equation}
   L=-m\left(1-\dot{a}^2-a^2 v^2 \right)^{1/2}
\label{eq:lagr}
\end{equation}
goes over into $-m+ma^2v^2/2$ in the limit of $av \ll 1$ and $\dot{a}=0$.
The motion of the particle is determined from the principle of least action,
$\delta S = -m\delta \int d s=0$ (e.g., $\S87$ of Ref. \cite{lan75}),
resulting in the geodesic equation
\begin{equation}
   \frac{d^2 x^i}{d s^2}
   +\Gamma^{i}_{kl} \frac{d x^k}{d s}\frac{d x^l}{d s} = 0.
\label{eq:geo}
\end{equation}

From the Lagrangian, we calculate energy and momentum of the material particle.
The 3-momentum of the particle is obtained from
$p_\alpha = \partial L / \partial v^\alpha$, with individual components
$p_1 = m\gamma v^1 a^2$, $p_2 = m\gamma v^2 a^2 \sin^2 \chi$,
and $p_3 = m\gamma v^3 a^2 \sin^2\chi \sin^2 \theta$.
The energy of the particle is given by 
\begin{equation}
   E_\textrm{p} = p_\alpha v^\alpha - L
     = {{m(1-\dot{a}^2)}\over{(1-\dot{a}^2 -a^2 v^2)^{1/2}}},
\label{eq:energy}
\end{equation}
and the relativistic mass by
\begin{equation}
   m_\textrm{r}=E_\textrm{p}/\eta^2 = m\gamma.
\label{eq:mr}
\end{equation}
Both energy and mass of a particle are tightly related to the expansion
speed of the universe. For a comoving particle with a fixed comoving
coordinate ($v^\alpha=0$), the energy and the relativistic mass become
$E_\textrm{p}=m (1-\dot{a}^2)^{1/2}$ and
$m_\textrm{r}=m (1-\dot{a}^2)^{-1/2}$, respectively. 

Let us define the 4-momentum vector of a particle as
\begin{equation}
   p^i = \left(m\gamma,m\gamma v\mathbf{n}\right)
       = \left({E_\textrm{p}\over{1-\dot{a}^2}}, {p\over a}\mathbf{n} \right),
\label{eq:momvec}
\end{equation}
where $p$ is the magnitude of the proper momentum defined as
$p = (-p^\alpha p_\alpha)^{1/2} = m \gamma a v$ and 
$\mathbf{n}$ is a unit vector indicating the direction of motion of 
the particle.
According to this definition, the particle's energy is the time component
of the covariant 4-momentum $p_k = mu_k = (E_\textrm{p},-ap\mathbf{n})$. 
From the square of the 4-momentum 
\begin{equation}
   p^i p_i = m^2, 
\end{equation}
we obtain an energy-momentum relation
\begin{equation}
   E_\textrm{p}^2 = \left( p^2 + m^2 \right) (1-\dot{a}^2),
\label{eq:ppm2}
\end{equation}
which goes over into $E_\textrm{p}^2 = p^2 + m^2$ in the limit of $\dot{a}=0$.
One important expectation from (\ref{eq:ppm2}) is that the energy of
a material particle vanishes when $\dot{a}=1$, e.g., at the beginning
of the universe.

The equation of motion of a particle with small peculiar velocity
($v^\alpha \ll 1$) is obtained from the space component of (\ref{eq:geo}) as
\begin{equation}
   \frac{d v^\alpha}{d t} + \left({{\dot{a}\ddot{a}}\over{1-\dot{a}^2}}
     +2{\dot{a}\over a}\right) v^\alpha = 0,
\label{eq:eqom}
\end{equation}
where any quadratic of $v^\alpha$ has been dropped.
The solution to this equation
\begin{equation}
   a v^\alpha \propto \frac{\sqrt{1-\dot{a}^2}}{a}
\end{equation}
shows how the proper peculiar velocity of a particle evolves as a result
of the cosmic expansion.

From (\ref{eq:momvec}) and (\ref{eq:ppm2}), the energy and the 4-momentum
of a massless photon are
\begin{equation}
   E_\gamma =p_\gamma (1-\dot{a}^2)^{1/2},
\label{eq:photon_e}
\end{equation}
and
\begin{equation}
   p^i = \left({p_\gamma\over\sqrt{1-\dot{a}^2}},{p_\gamma\over a}
         \mathbf{n}\right),
\end{equation}
where $p_\gamma$ is the photon's proper spatial momentum.
The photon's energy and spatial momentum are usually expressed as
photon's frequency and inverse wavelength multiplied by the Planck
constant ($E_\gamma = h \nu$ and $p_\gamma = h / \lambda$). 
Therefore, Eq. (\ref{eq:photon_e}) is equivalent to
\begin{equation}
   \nu\lambda = \eta = (1-\dot{a}^2)^{1/2}.
\label{eq:nulam}
\end{equation} 
A comoving observer measures frequency and wavelength of the same photon as
$\nu_\textrm{c} = \nu (1-\dot{a}^2)^{-1/2}$ and $\lambda_\textrm{c} =\lambda$.
Thus, $\nu_\textrm{c} \lambda_\textrm{c} = 1$ in the locally inertial frame.
The subscript $\textrm{c}$ denotes a quantity measured by the comoving observer.

\subsection{Doppler shift and cosmological redshift of photons}
\label{sec:doppz}

The stretch of photon's wavelength is induced by the receding motion
of an observer from a light source (Doppler shift), or by the cosmic
expansion (cosmological redshift).

First, let us consider the Doppler shift.
For simplicity, the cosmic expansion speed is assumed to be fixed.
Suppose that an observer with 4-velocity
$u^{i}=(\gamma,\gamma v \mathbf{n}_1)$
is moving away from a light source emitting photons with 4-momentum 
$p^{i}=({p_\textrm{em}/\sqrt{1-\dot{a}^2}},p_\textrm{em} 
{\mathbf{n}_2}/a)$, and is receiving photons from the source.
The observed momentum of a photon is given by the inner product
of $u^i$ and $p^i$:
\begin{equation}
   p_\textrm{ob} = p^i u_i 
      = {{\sqrt{1-\dot{a}^2}-av\cos\theta_{12}}
        \over{\sqrt{1-\dot{a}^2-a^2 v^2}}} p_\textrm{em},
\end{equation}
where $p_\textrm{em}$ and $p_\textrm{ob}$ are the proper spatial momenta 
of emitted and observed photons, respectively, and 
$\mathbf{n}_1\cdot\mathbf{n}_2=\cos\theta_{12}$ ($\theta_{12}=0$
for receding and $\theta_{12}=\pi$ for approaching observers).
The ratio of momenta (or energies) of observed to emitted photons
for the longitudinal Doppler effect ($\theta_{12}=0$) is
\begin{equation}
   {{p_\textrm{ob}}\over{p_\textrm{em}}}
   = {{E_\textrm{ob}}\over{E_\textrm{em}}}
    =\sqrt{{c_\chi-v}\over{c_\chi+v}},
\label{eq:doppler1}
\end{equation}
where $c_\chi \equiv \eta / a$ is the light speed in the comoving
coordinate system.
The ratio for the transversal Doppler effect ($\theta_{12}=\pi/2$) is
\begin{equation}
   {{p_\textrm{ob}}\over{p_\textrm{em}}}
   = {{E_\textrm{ob}}\over{E_\textrm{em}}}
   = {1\over\sqrt{1-v^2/c_\chi^2}}.
\label{eq:doppler2}
\end{equation}
The two formulas for the Doppler effect are similar in form
to those in the special relativity.

Next, for the cosmological redshift, let us suppose that photons,
emitted at world time $t$ from a light source at comoving coordinate $\chi$, 
have arrived at the origin at $t_0$. 
Using the photon's geodesic equation and assuming that photons
have traveled radially by the symmetry of space, we get 
\begin{equation}
\begin{split}
 \int_{0}^{\chi} d\chi' 
         &= -\int_{t_0}^{t} 
                   {{\sqrt{1-\dot{a}^2}}\over{a}} d t'  \\
         &= \int_{t}^{t_0} c_\chi(t') d t' 
              = \int_{t+\delta t}^{t_0 +\delta t_0} c_\chi(t') d t',
\end{split}
\label{eq:redshift}
\end{equation}
where the minus sign in front of the second integral indicates
that photons have propagated from the distant source to the origin.
The $\delta t$ and $\delta t_0$ are the world time intervals
during which a photon's wave crest propagates
by the amount of its wavelength at the points of emission and observation,
respectively (i.e., $\nu=\delta t^{-1}$).
The third equality holds because the integral does not change 
after the infinitesimal time intervals $\delta t$ and $\delta t_0$.

\begin{figure}
\mbox{\epsfig{file=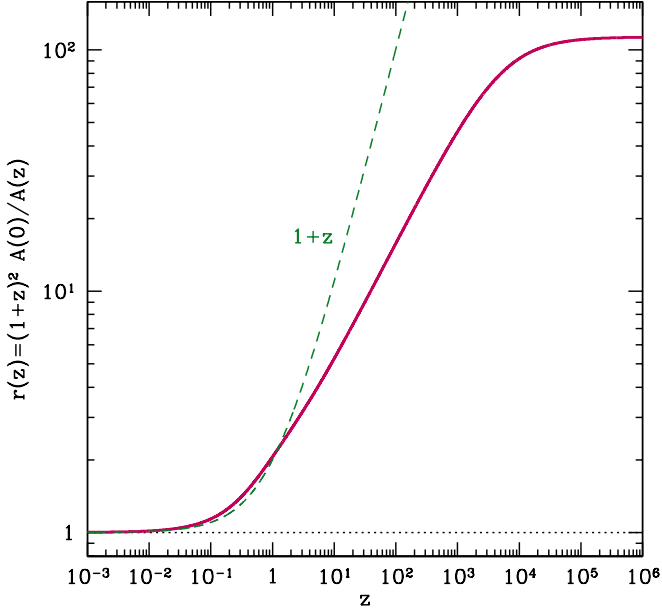,width=88mm,clip=}}
\caption{Variation of $r(z)$ along with redshift (thick solid curve; Model I).
         The upper bound value of $r(z)$ at infinite redshift is $113$.
         For comparison, the cosmological time dilation factor $1+z$
         in the FRW world model is shown as a dashed curve.}
\label{fig:dt0dtz}
\end{figure}

Manipulating (\ref{eq:redshift}) gives the frequency ratio of
emitted to observed photons,
\begin{equation}
\begin{split}
   {{\nu} \over {\nu_0}} 
    &= {{\delta t_0}\over{\delta t}} 
    = {{c_\chi(t)}\over{c_\chi(t_0)}}  \\
    &= {a_0\over a}\left({{1-\dot{a}^2}\over{1-\dot{a}_0^2}}\right)^{1/2} 
    = (1+z)^2 {{A(0)}\over{A(z)}} \equiv r(z).
\end{split}
\label{eq:photon_rate}
\end{equation}
The cosmological time dilation function $r(z)$ is useful for comparing
physical quantities at past and present epochs.
The variation of $r(z)$ is shown in Fig. \ref{fig:dt0dtz}.
The corresponding function in the FRW world model, $1+z$,
has a similar value to $r(z)$ only at low redshift ($z \la 1$),
going to infinity at infinite redshift.
The $r(z)$ is almost constant during the radiation-dominated era
with the maximum value of
\begin{equation}
   r(\infty)
     =[(\varepsilon_{\textrm{R}0}+\varepsilon_{\textrm{M}0}+\varepsilon_\Lambda)
      /\varepsilon_{\textrm{R}0}]^{1/2}=113. 
\label{eq:rz}
\end{equation}

From (\ref{eq:nulam}) and (\ref{eq:photon_rate}), the photon's frequency
and wavelength vary as 
\begin{equation}
   \nu \propto \frac{\sqrt{1-\dot{a}^2}}{a} \mskip+12mu \textrm{and}
       \mskip+12mu \lambda \propto a.
\end{equation}
Because the wavelength of a photon increases in proportional to $a$,
the redshift is equal to the fractional difference between wavelengths 
at the points of observation and emission of the photon: $z = a_0/a -1 
= (\lambda_\textrm{ob}-\lambda_\textrm{em})/\lambda_\textrm{em}$.
Besides, the photon's energy and spatial momentum vary as
\begin{equation}
  E_\gamma \propto \frac{\sqrt{1-\dot{a}^2}}{a} \mskip+12mu \textrm{and} 
     \mskip+12mu p_\gamma \propto \frac{1}{a}.
\label{eq:phot_em}
\end{equation}
From (\ref{eq:rde}), (\ref{eq:mde}), and (\ref{eq:dde}),
one finds that $E_\gamma = \textrm{const.}$ (R-u), 
$E_\gamma \propto a^{-1/2}$ (M-u), and
$E_\gamma \propto a^{-2}$ ($\Lambda$-u).
In the comoving observer's frame, both photon's energy and spatial momentum
always vary as $a^{-1}$.

\subsection{Total energy in the universe}
\label{sec:energy}

The conservation of energy in the classical physics is closely related
to the invariance of physical laws under a time-translation
(Noether's theorem), which applies to the physics in the Minkowski space-time.
In general relativity there is not necessarily a time coordinate
with the translation-symmetry, so the conservation of energy is not
generally expected. However, in an asymptotically flat region or in a locally
inertial frame, it is possible to define the conserved energy.
For this reason, it has been usually said that there is not a global
but a local energy conservation law.

Let us estimate the total energy in the universe based on the definition
of energy in Sec. \ref{sec:energy_momentum}. 
First, we need to define the volume element.
The 4-dimensional volume element is given by
\begin{equation}
   d V_4 = \sqrt{-g} d x^0 d x^1 d x^2 d x^3
            = \sqrt{-g} d t d\chi d\theta d\phi,
\end{equation}
where $g$ is the determinant of the metric tensor $g_{ik}$.
We obtain the 4-volume of the universe from
$V_4 (t) = \int d V_4 = 2\pi^2 \int_{0}^{t} (1-\dot{a}^2)^{1/2} a^3 d t'$.
The proper 3-volume of the universe is the time-derivative of $V_4$:
\begin{equation} 
   V(t) = \frac{d V_4}{d t} = 2\pi^2 (1-\dot{a}^2)^{1/2} a^3 
       = \frac{2\pi^2 a_0^3}{(1+z)^2 A(z)}.
\label{eq:vol3}
\end{equation}
The factor $(1-\dot{a}^2)^{1/2}$ appears as a natural contraction
effect due to the expansion of the universe.
At the present time, $V_0 = 6.94\times 10^{13}$ Mpc$^3$.
One can verify that the 3-volume of the universe evolves
as $V\propto a^4$ (R-u), $V\propto a^{7/2}$ (M-u), and $V\propto a^2$
($\Lambda$-u).
Note that $V_\textrm{c} = d V_4 /d\tau \propto a^3$ in the comoving
observer's frame and thus $V=(1-\dot{a}^2)^{1/2} V_\textrm{c}$.

If there are $N$ comoving particles with rest mass $m$ in the universe,
then the matter energy density is 
\begin{equation}
   \varepsilon_\textrm{M} 
   = {{N E_\textrm{p}}\over{V}} 
   = {{N m (1-\dot{a}^2)^{1/2}} \over{(1-\dot{a}^2)^{1/2} V_\textrm{c}}} 
   = {{N m}\over{V_\textrm{c}}} = \varepsilon_\textrm{Mc}.
\label{eq:medensity}
\end{equation}
Therefore, matter energy densities both in the world reference
and the comoving observer's frames are equivalent to each other,
which also applies to the radiation energy density if $m$ is replaced
with $p_\gamma$ in (\ref{eq:medensity}).
The matter energy density is related to the matter density $\rho_\textrm{M}$
by $\varepsilon_\textrm{M} = \rho_\textrm{M}\eta^2$ because
$\rho_\textrm{M} = N m_\textrm{r}/V = N m / V_\textrm{c} (1-\dot{a}^2)
= \varepsilon_\textrm{M} / (1-\dot{a}^2)$.

\begin{figure}
\mbox{\epsfig{file=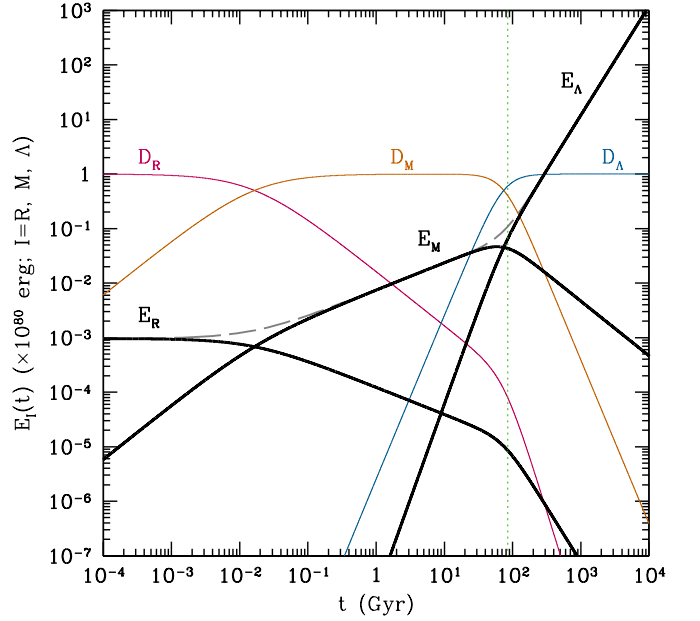,width=88mm,clip=}}
\caption{Variation of total radiation ($E_\textrm{R}$),
         matter ($E_\textrm{M}$), and dark ($E_{\Lambda}$)
         energies in unit of $10^{80}$ erg over the world time $t$
         (thick solid curves; Model I), together with that of total energy
         $E=E_\textrm{R} + E_\textrm{M} + E_\Lambda$ (long dashed curve).
         Thin solid curves represent the variation of the corresponding
         energy fraction parameters ($D_\textrm{R}$, $D_\textrm{M}$, $D_\Lambda$;
         $D_\textrm{R}+D_\textrm{M}+D_\Lambda=1$) with a dimensionless unit.
         The vertical dotted line indicates $t_0$.}
\label{fig:energy_plot}
\end{figure}

The total energy of each species $I$ is calculated from
\begin{equation}
   E_{I}=\varepsilon_{I} V
        =2\pi^2 a_0^3 \varepsilon_{I0} \frac{(1+z)^{1+3w_I}}{A(z)}.
\end{equation}
The present values of total radiation, matter, and dark energies 
($E_{\textrm{R}0}$, $E_{\textrm{M}0}$, $E_{\Lambda0}$) are listed
in Table \ref{tab:cospar}.
In radiation, matter, and $\Lambda$-universes, the total energy evolves as
$E_\textrm{R}=\textrm{const.}$, $E_\textrm{M}\propto a^{1/2}$, and
$E_\Lambda \propto a^2$, respectively.
On the other hand, $E_\textrm{Rc} \propto a^{-1}$, $E_\textrm{Mc}=\textrm{const.}$,
and $E_{\Lambda\textrm{c}}\propto a^3$ in the comoving observer's frame,
implying that the radiation energy is infinite at the initial time,
and the matter energy is always constant.

Fig. \ref{fig:energy_plot} shows the variation of total radiation, matter,
and dark energies during the history of the universe (thick solid curves),
together with that of energy fraction parameters ($D_I$; thin solid curves).
The total radiation energy remained constant in the radiation-dominated era
($t \la 10^{-3}$ Gyr), and thereafter has decreased.
The total matter energy was zero at $t=0$, arrived at the maximum value
at $t=58.8$ Gyr ($\tau = 10.0$ Gyr), and then has decreased,
going back to zero far in the future.
The total dark energy has continuously risen up from zero
as the universe expands.
We can estimate the total rest mass in the universe by transforming
the total matter energy into 
$M = E_{\textrm{M}0}(1-\dot{a}_0^2)^{-1/2} \approx 10^{25} M_\odot$,
which corresponds to about $10^{14}$ galaxies with a typical mass
of $10^{11} M_\odot$.

From the definition of energy fraction parameters 
[Eqs. (\ref{eq:di}) and (\ref{eq:ddd})], one finds that
$\varepsilon V c_\chi$ is a conserved quantity such that
\begin{equation}
   \varepsilon(t) V(t) c_\chi(t) 
     = \varepsilon_0 V_0 c_{\chi 0}=\textrm{const.},
\end{equation}
where $\varepsilon$ is the total energy density.
Thus, the ratio of present to past total energies in the universe
is obtained as 
\begin{equation}
   \frac{E_0}{E(t)} 
    = { {\varepsilon_0 V_0 }
        \over{\varepsilon(t) V(t) } }
    =   {{c_\chi (t) }\over{c_\chi (t_0) }} = r(z),
\end{equation}
where $E=\sum_I E_I$ (long dashed curve in Fig. \ref{fig:energy_plot})
and $E_0 = 1.09 \times 10^{79}$ erg.
The initial amount of total energy in the universe is
$E(0) = E_0 / r(\infty) = 9.62 \times 10^{76}$ erg.
Since only the radiation contributes to the total energy at $t=0$,
the same value is obtained from
$E(0) = E_\textrm{R} (0) = E_{\textrm{R}0} r(\infty)$
with the help of $E_\textrm{R} (t) = E_{\textrm{R}0} r(z)$ deduced from 
(\ref{eq:photon_rate}).

According to the definition of energy in this paper, 
the total energy is not conserved in the expanding closed universe,
but increases with time.
Especially, the total energy is finite at the beginning of the universe.

\subsection{Energy density, pressure, and temperature in thermal equilibrium}
\label{sec:relgas}

We describe the evolution of energy density, pressure, and temperature
of gas in the early universe (see \cite{kolb90} for details).
Our discussion is restricted to the relativistic gas particles in thermal
equilibrium. The particles are assumed to have low rest mass compared to
their kinetic energy.

The number of particles of species $I$ per unit spatial volume $d V$
per unit momentum volume $d W$ can be expressed as
\begin{equation}
   d N_I = {{g_I f_I (x^i,p_i)}\over{(2\pi\hbar)^3}} d V d W,
\end{equation}
where $g_I$ is the spin-degeneracy of the particle,
$f_I$ is the particle distribution function for species $I$, equivalent
to the mean number of particles occupying a given quantum state,
and $\hbar=h/2\pi$.
Assuming zero chemical potential, we can write 
$f_I = [\exp(E_\textrm{p}/k_\textrm{B} T_I)\pm 1]^{-1}$, where plus and minus signs
are for fermions ($\textrm{f}$) and bosons ($\textrm{b}$), respectively,
$k_\textrm{B}$ is the Boltzmann constant, and $T_I$ is the thermodynamic
temperature of species $I$.
The proper spatial and momentum volume elements in the world
reference frame are given by
\begin{eqnarray}
   d V &=& \sqrt{-g} d\chi d\theta d\phi 
                  = (1-\dot{a}^2)^{1/2} d V_\textrm{c}, \nonumber \\
   d W &=& {1\over\sqrt{-g}} d p_\chi d p_\theta d p_\phi 
                  = (1-\dot{a}^2)^{-1/2} d W_\textrm{c},
\end{eqnarray}
where $d V_\textrm{c}$ and $d W_\textrm{c}$ are proper volume elements
in the comoving observer's frame.
Note that $d V d W = d V_\textrm{c} d W_\textrm{c}$.
The momentum volume element is written as
$d W = (1-\dot{a}^2)^{-1/2} 4\pi p^2 d p$ for isotropic gas particles
with proper momentum $p$.

The energy density of the relativistic gas is obtained
by integrating over the momentum space the particle's energy multiplied
with its distribution function:
\begin{equation}
\begin{split}
   \varepsilon_I 
      &= {g_I\over(2\pi\hbar)^3} \int E_\textrm{p} f_I d W
      = {g_I\over(2\pi\hbar)^3} \int_0^\infty 
        {{4\pi p^3 d p}\over{e^{E_\textrm{p}/k_\textrm{B} T_I}\pm 1}} \\
      &= {{4\pi g_I k_\textrm{B}^4 T_I^4}\over{(2\pi\hbar)^3} (1-\dot{a}^2)^{2}}
             \int_0^\infty {{x^3}\over{e^x \pm 1}} d x  \\
      &= \frac{7\pi^2 g_I k_\textrm{B}^4 T_I^4}{240\hbar^3
          (1-\dot{a}^2)^2} \mskip+12mu (\textrm{f}) \mskip+12mu
          \textrm{and} \mskip+12mu
         \frac{\pi^2 g_I k_\textrm{B}^4 T_I^4}{30\hbar^3
          (1-\dot{a}^2)^2} \mskip+12mu (\textrm{b}),
\end{split}
\label{eq:eden}
\end{equation}
where $x \equiv E_\textrm{p}/k_\textrm{B} T_I$ and
$E_\textrm{p}\simeq p (1-\dot{a}^2)^{1/2}$ for the relativistic gas. 

The pressure of the relativistic gas is obtained in a similar way:
\begin{equation}
\begin{split}
   P_I &= {g_I\over(2\pi\hbar)^3}\int p_\textrm{1d} v_\textrm{1d} f_I d W \\
       &= {g_I\over{3(2\pi\hbar)^3}} \int_0^\infty 
         {{4\pi p^3 d p}\over{e^{E_\textrm{p}/k_\textrm{B} T_I}\pm 1}} 
       = {\varepsilon_I \over 3},
\end{split}
\label{eq:epre}
\end{equation}
where $p_\textrm{1d}$ and $v_\textrm{1d}$ are proper momentum
and velocity in one direction:
$v_\textrm{1d} = (p_\textrm{1d}/E_\textrm{p})(1-\dot{a}^2)$
and $p_\textrm{1d}^2 = p^2 /3$ for isotropic gas. 
The relativistic gas acts like radiation, with an equation of state
$w_I = \frac{1}{3}$ and the energy density varying as
$\varepsilon_I \propto a^{-4}$.
Therefore, from (\ref{eq:eden}) the thermodynamic temperature of
the relativistic gas evolves as 
\begin{equation}
   T_I \propto {{\sqrt{1-\dot{a}^2}}\over{a}}.
\label{eq:ttrg}
\end{equation}
In the comoving observer's frame, $T_{I\textrm{c}}\propto a^{-1}$.
From a formula of the entropy density
$\sigma_I=(\varepsilon_I+P_I)/k_\textrm{B} T_I$,
one can verifies that the total entropy $S_I=\sigma_I V$ of
the relativistic gas is constant. 

For photons, the quantity $x=h\nu/k_\textrm{B} T_\textrm{R}$ is invariant
during the cosmic expansion history because the photon's frequency varies  
in the same way as the temperature does.
Since $x$ is also frame-independent ($x=x_\textrm{c}$), we have
$T_\textrm{R}=T_\textrm{Rc} (\nu/\nu_\textrm{c})=T_\textrm{Rc} (1-\dot{a}^2)^{1/2}$.
The present CMB temperature in the world reference frame is
$T_{\textrm{R}0} = T_{\textrm{Rc}0} (1-\dot{a}_0^2)^{1/2}=0.57~\textrm{K}$
($T_{\textrm{Rc}0}=T_\textrm{cmb}$).
From (\ref{eq:ttrg}), the ratio of past to present radiation temperatures is 
\begin{equation}
   \frac{T_\textrm{R}}{T_{\textrm{R}0}} 
       = \frac{a_0}{a}
         \left(\frac{1-\dot{a}^2}{1-\dot{a_0}^2}\right)^{1/2} = r(z),
\end{equation}
which enables us to estimate the radiation temperature at the past epoch.
For example, at the beginning of the universe 
$T_\textrm{R} (0) = T_{\textrm{R}0} r(\infty) = 64.0~\textrm{K}$, while it is infinite
in the comoving observer's frame.
The behavior of $r(z)$ implies that $T_\textrm{R} = \textrm{const.}$
in the radiation-universe.

The epoch of radiation-matter equality is determined from
the condition $\varepsilon_\textrm{M} = \varepsilon_\textrm{R}$: 
\begin{equation} 
  1+z_\textrm{eq} = \frac{\varepsilon_{\textrm{M}0}}{\varepsilon_{\textrm{R}0}} 
          = \left({{a_{\textrm{R}0}}\over{a_{\textrm{M}0}}}\right)^2 = 5081.
\end{equation}
At this epoch ($t_\textrm{eq}=1.65\times 10^{-2}$ Gyr,
$\tau_\textrm{eq}=29700$ yr), the size of the universe was
$a_\textrm{eq} = 5.05$ Mpc and the radiation temperature was
$T_\textrm{R, eq} = T_{\textrm{R}0} r(z_\textrm{eq}) = 45.3$ K, or
$T_\textrm{Rc, eq} = T_{\textrm{Rc}0} (1+z_\textrm{eq}) = 13850$ K.

To summarize, as judged in the world reference frame,
the early universe was cold and all the physical processes in it
were extremely slow.
Especially, the universe started from a regular (non-singular) point
in the sense that physical quantities have finite values at the initial time. 
The singular nature of the FRW universe comes from the fact
that the flow of the proper time was frozen ($d\tau=0$) at $t=0$.

\subsection{Cosmic distance and time scales}
\label{sec:cosmic_scales}

Lastly, we consider cosmic distance and time scales 
in the closed world model.
As the most popular distance measure, the coordinate distance ($d_\textrm{C}$)
to a galaxy at redshift $z$ is obtained by integrating the photon's
geodesic equation ($d\chi = d t \sqrt{1-\dot{a}^2}/a$),
\begin{equation}
\begin{split}
    d_\textrm{C} (z) &= a_0 \chi_\textrm{C} (z) 
        = a_0 \int_{t}^{t_0} {{\sqrt{1-\dot{a}^2}}\over{a}} d t' \\
        &= \int_0^z {{a_0 d z'}\over{\left[ A^2(z')-(1+z')^2 \right]^{1/2}}}.
\end{split}
\label{eq:dc}
\end{equation}
Here $\chi_\textrm{C} (z)$ is the comoving coordinate distance.
For sufficiently large $a_0$, Eq. (\ref{eq:dc}) goes over into
$d_\textrm{C} (z) \approx \int_0^z a_0 d z'/A(z')$, which is equivalent to 
the coordinate distance in the flat FRW world model.
The coordinate distance in the matter-universe has an analytic form
\begin{equation}
\begin{split}
   d_\textrm{C}(z) &= 2a_0
      \arctan\left[\left(a_0\over a_{\textrm{M}0}\right)^2 (1+z)
                  -1\right]^{1/2} \\
     &-2a_0 \arctan\left[\left(a_0\over a_{\textrm{M}0}\right)^2
                  -1 \right]^{1/2} \mskip+24mu (\textrm{M-u}).
\end{split}
\end{equation}

We can derive other astronomical distances based on luminosity and
angular size of distant sources. 
For the luminosity distance, let us imagine that a light source
at redshift $z$ has intrinsic bolometric luminosity $L_\textrm{c}$ as measured
at the source.
Since both photon's energy and arrival rate vary in proportion to $a^{-1}$
in the locally inertial frame, the flux of the light source
as measured by the present comoving observer can be written as
\begin{equation}
  {L_\textrm{c}\over{4\pi d_\textrm{L}^2}}
     = {L_\textrm{c}\over{4\pi a_0^2 \sin^2\chi_\textrm{C}}}
     \left(a\over a_0\right)^2,
\end{equation}
where $d_\textrm{L}$ is the luminosity distance to the source
\footnote{If the flux of the source is measured in the world reference
frame, the luminosity distance becomes
$d_\textrm{L} (z) = a_0 r(z) \sin\chi_\textrm{C} (z)$.},
\begin{equation}
   d_\textrm{L}(z) = a_0 (1+z) \sin\chi_\textrm{C} (z).
\label{eq:ldist}
\end{equation}
The angular size distance ($d_\textrm{A}$) to a galaxy with physical size
$r_\textrm{g}$ and angular size $\theta_\textrm{g}$ is given by
\begin{equation}
   d_\textrm{A} (z) = {r_\textrm{g}\over\theta_\textrm{g}} 
         = {\sqrt{g_{22}}\theta_\textrm{g}\over\theta_\textrm{g}}
         = {a_0\sin\chi_\textrm{C}(z)\over{1+z}}.
\end{equation}

The recession velocity ($v_\textrm{rec} = z$) of a galaxy has also been
used as a distance measure in the local universe.
By integrating $d z=-d a (a_0/a^2)$ from the definition of redshift,
we get
\begin{equation}
   z = \int_t^{t_0}\left(\frac{a_0}{a}\right)
       \left(\frac{\dot{a}}{a}\right)d t'
      = a_0 \int_t^{t_0} H(\tau') \frac{\sqrt{1-\dot{a}^2}}{a} d t',
\end{equation}
where $H(\tau)=a^{-1} (d a/d\tau)=a_0^{-1} [A^2 (z)-(1+z)^2 ]^{1/2}$
is the Hubble parameter.
Since the Hubble parameter remains almost constant during the recent epoch 
($z\la 0.1$), the recession velocity is related to
the coordinate distance by $v_\textrm{rec} \approx H_0 d_\textrm{C} (z)$,
which is the Hubble's law.

The age of the universe or the lookback time have been used as a measure
of cosmic time scales.
The age of the universe measured in world time is calculated from
\begin{equation}
  t(z) = \int_{0}^{t} d t'
       = \int_z^\infty {{a_0 d z'}\over{(1+z')^2}} 
         \sqrt{{A^2(z')}\over{A^2(z')-(1+z')^2}},
\label{eq:agew}
\end{equation}
while the age measured in proper time of a comoving observer (us)
is obtained by integrating (\ref{eq:tau}):
\begin{equation}
\begin{split}
   \tau(z) &= \int_0^{t} \sqrt{1-\dot{a}^2} d t' \\
           &= \int_z^\infty \frac{a_0 d z'}{(1+z')}
                \frac{1}{\left[A^2(z') - (1+z')^2\right]^{1/2}},
\end{split}
\label{eq:agec}
\end{equation}
The latter is equivalent to the age of the FRW universe.
Note that $t\ge\tau$ (see Fig. \ref{fig:tplot} for $\tau$-$t$ relation).
At the present time, $t_0=85.3$ Gyr and $\tau_0=15.8$ Gyr.
The lookback time, the time measured back from the present to the past,
is given by $t_0 - t(z)$ or $\tau_0 - \tau(z)$.

\begin{figure}
\mbox{\epsfig{file=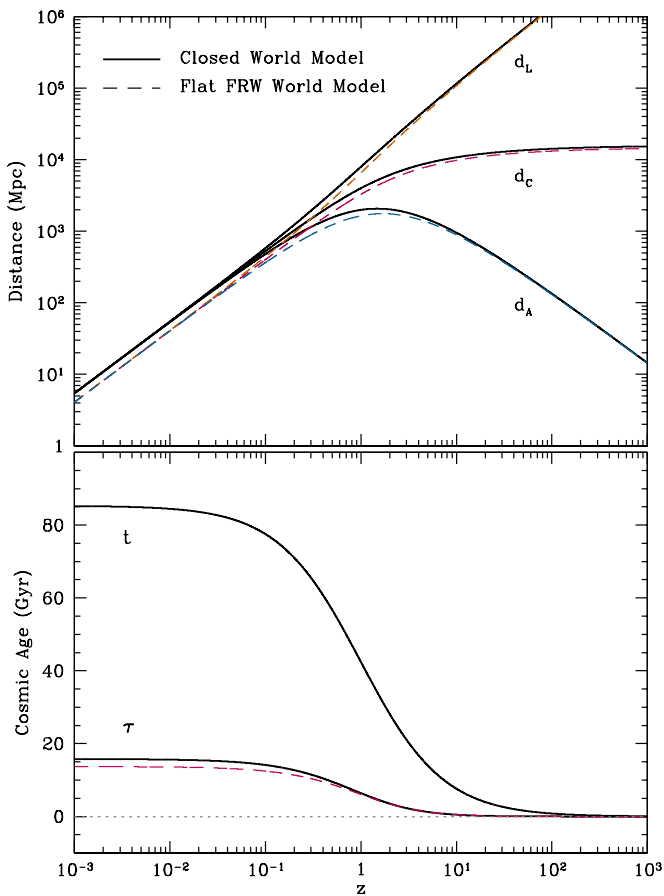,width=88mm,clip=}}
\caption{(top) Coordinate ($d_\textrm{C}$), luminosity ($d_\textrm{L}$),
         and angular size ($d_\textrm{A}$) distances in the closed world
         model (solid curves; Model I).
         The coordinate distance to the big-bang is
         $d_\textrm{C} (\infty) = 15.7$ Gpc.
         (bottom) Cosmic ages measured in world time ($t$; top curve)
         and in proper time of a comoving observer ($\tau$; bottom curve).
         The dashed curves are the corresponding distances and age
         for the flat FRW world model that best fits with the WMAP CMB data
         ($\Omega_\textrm{M} h^2 = 0.1277$ and $h=0.732$; \cite{spergel07}). }
\label{fig:dist}
\end{figure} 

Fig. \ref{fig:dist} compares coordinate, luminosity, angular size distances 
(top) and cosmic ages (bottom) as a function of redshift in the closed
world model with Model I parameters of Table \ref{tab:cospar}.
Also shown are the corresponding distances and age for the flat FRW
world model that best fits with the WMAP CMB data
($\Omega_\textrm{M} h^2 = 0.1277$ and $h=0.732$; \cite{spergel07}).
Distances of both world models agree with each other at high redshift
($z \ga 10$), but the flat FRW world model underestimates distances
to nearby galaxies than the closed world model.

\section{Inflation}
\label{sec:inflation}

The FRW world model has been criticized because of two shortcomings,
namely, flatness and horizon problems. 
As shown in Sec. \ref{sec:nspace}, the universe with positive energy
density is always spatially closed.
If the present universe is traced back to the past, it would become
a more curved hypersurface, a 3-sphere with smaller curvature radius.
Therefore, the closed world model proposed in this paper is free from
the flatness problem.
Next, let us consider the horizon problem, which is stated as follows.
The observed CMB temperature fluctuations separated by more than a degree
are similar to each other over the whole sky.
In the FRW world model, such an angle corresponds to a distance
where the causal contact was impossible on the last scattering surface.
The large-scale uniformity of the CMB anisotropy suggests that
the observed regions must have been in causal contact in the past.

The inflation paradigm has offered a reasonable solution to 
the puzzle of the large-scale homogeneity of the observable universe
by proposing that there was a period of rapid expansion of the universe
with positive acceleration \cite{guth81}.
According to the inflation theory, the inflation takes place due to
the presence of a scalar field $\phi$, whose energy density
and pressure are given by
$\varepsilon_\phi = \frac{1}{2} {\dot{\phi}}^2 + V(\phi)$ and
$P_\phi = \frac{1}{2} {\dot{\phi}}^2 - V(\phi)$, respectively,
where $V(\phi)$ is a potential of the scalar field. 
Here we assume that the dot over $\phi$ is the world-time derivative.
From (\ref{eq:evol_closed1}), for a universe dominated by
the scalar field, the condition for the positive expansion acceleration is
\begin{equation}
   \dot{\phi}^2  < V(\phi).
\label{eq:inf_cond}
\end{equation}

A distance scale of causally connected region (so called horizon size)
is usually quantified by the Hubble radius and the particle horizon size.
In the FRW world model, one important implication of the positive
expansion acceleration is that the comoving Hubble radius decreases
with time, i.e., $d (aH)^{-1} /d\tau < 0$.
The comoving Hubble radius is defined as the comoving distance
at which the recession velocity as defined in the world reference frame
is equal to the speed of light ($\dot{a} \chi_\textrm{H} = \eta$):
\begin{equation}
   \chi_\textrm{H} (t) = \frac{\sqrt{1-\dot{a}^2}}{\dot{a}}
          = \frac{1}{aH(\tau)}
          =  {{1+z}\over{\left[A^2 (z) - (1+z)^2 \right]^{1/2}}}. 
\end{equation}
The comoving particle horizon size is the comoving distance a photon
has travelled during the age of the universe:
\begin{equation}
   \chi_\textrm{P} (t) = \int_0^t {\sqrt{1-\dot{a}^2}\over a} d t' 
              = \int_z^\infty 
             {{d z'}\over{\left[A^2 (z') - (1+z')^2\right]^{1/2}}}.
\end{equation}
The proper Hubble radius and particle horizon size are given by
$d_\textrm{H} (t) = a(t) \chi_\textrm{H} (t) = 1/ H(\tau)$ and
$d_\textrm{P} (t) = a(t) \chi_\textrm{P} (t)$,
respectively.

\begin{figure}
\mbox{\epsfig{file=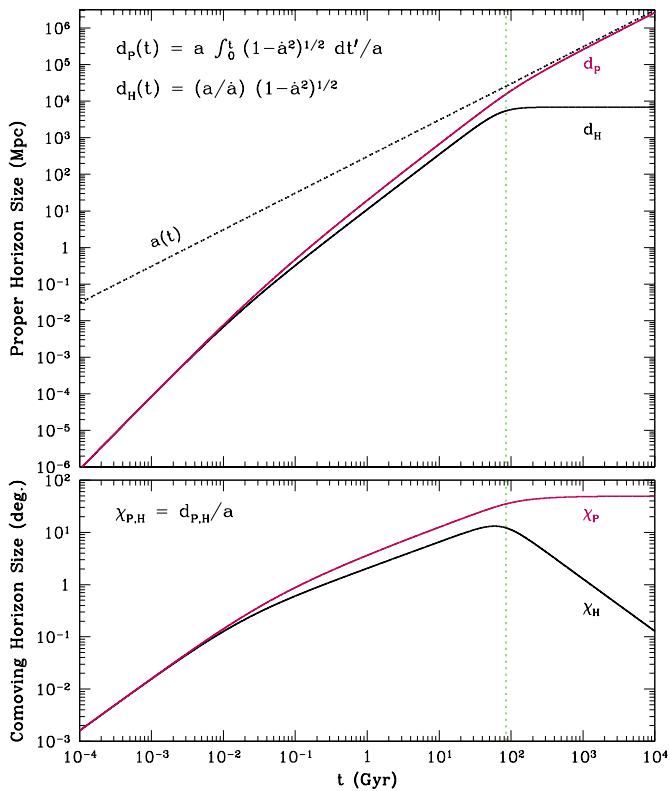,width=88mm,clip=}}
\caption{Time-variation of proper (top) and comoving (bottom) horizon sizes
         in the closed universe (Model I), namely, the Hubble radius
         ($d_\textrm{H}$, $\chi_\textrm{H}$) and the particle horizon size
         ($d_\textrm{P}$, $\chi_\textrm{P}$).
         For comparison, the evolution of curvature radius
         $a(t)$ is shown as dashed curve in the top panel.
         The proper Hubble radius has an asymptotic maximum
         $d_\textrm{H} (\infty) = a_\Lambda = 6867$ Mpc.
         The comoving horizon size is shown in unit of degree,
         with an asymptotic maximum $\chi_\textrm{P} (\infty) = 49\fdg4$.
         The vertical dotted lines indicate the present epoch.}
\label{fig:horizon}
\end{figure}

Fig. \ref{fig:horizon} compares proper (top) and comoving (bottom)
horizon sizes as a function of world time.
It is important to note that comoving horizon sizes were
zeros at the beginning of the universe, and then have increased
until the recent epoch.
As shown in Sec. \ref{sec:evhist}, the universe was expanding with
the limiting speed and zero acceleration at the initial time
(Figs. \ref{fig:adotplot} and \ref{fig:accplot}).
Therefore, the positive expansion acceleration or the decrease
in the comoving Hubble radius is not allowable at the early stage
of the closed universe. The decrease is only possible
in later $\Lambda$-dominated universe ($t\ga 100$ Gyr;
Fig. \ref{fig:horizon}, bottom).

Actually, the scalar field $\phi$ is not essential for driving the rapid
expansion of the universe. 
Even if the scalar field is dominant, the condition (\ref{eq:inf_cond})
is not satisfied. We can only expect that $\dot{\phi}^2 = V(\phi)$
from the constraint of zero acceleration, obtaining an equation of state
\begin{equation}
   P_\phi = -\frac{\varepsilon_\phi}{3}.
\end{equation}
Thus the curvature radius increases as $a\propto t$ in the universe
dominated by the scalar field. Further constraining the universe
to expand with the limiting speed demands that the energy density
of the scalar field should be infinite, as in the extreme case of H-universe
(Sec. \ref{sec:evhist}).
However, the radiation-universe with finite total energy provides
a far simpler expansion history $a\simeq t$ for $t \ll b_\textrm{R}$
[Eq. (\ref{eq:rde})],
which demonstrates the sufficiency of the radiation in driving the rapid
expansion and the needlessness of the scalar field.
In conclusion, it is improbable that the inflation with positive
acceleration occurred in the early universe.

If the universe expands with the limiting speed,
the peculiar velocity of a particle vanishes as implied by (\ref{eq:dtau}):
the matter and radiation were frozen with zero propagation speed
at the earliest epoch. Besides, information at one region could not
be easily transferred to other regions due to the small horizon size.
Thus, physical information sharing through the causal contact during
the expansion of the universe is not an efficient way to explain for
the large-scale homogeneity of the universe.
The spherically symmetric and uniform distribution of supernova remnants
(e.g., Tycho's supernova 1572; \cite{warren05})
driven by a strong shock into the ambient interstellar medium
shows that the large-scale homogeneity can be generated from the ballistic
explosion at the single point, without the causal contact during the expansion.
At the beginning of the universe, everything was on the single point
so that every information such as temperature and energy density
could be shared in full and uniform contact.
Therefore, if the initial condition was properly set at the creation
of the universe, e.g., by the quantum processes at $t \la t_\textrm{p}$
(Planck time), which is out of the scope of the classical physics,
the observed uniformity of density distributions at super-horizon scales
may be explained.

\section{Conclusion}
\label{sec:conc}

In this paper, the general world model for homogeneous and isotropic
universe has been proposed. By introducing the world reference frame
as a global and fiducial system of reference, we have defined the line
element so that the effect of cosmic expansion on the physical space-time
separation can be correctly included in the metric.
With this framework, we have demonstrated theoretically that
the flat universe is equivalent to the Minkowski space-time
and that the universe with positive energy density is always spatially
closed and finite.
The open universe is unrealistic because it cannot accommodate positive
energy density. Therefore, in the world of ordinary materials,
only the spatially closed universe is possible to exist.

The naturalness of the finite world with positive energy density
comes from the Mach's principle that the motion of a mass particle
depends on the mass distribution of the entire world.
The principle is consistent only with the finite world because the dynamics
of a reference frame cannot be defined in the infinite, empty world.
The closed world model satisfies the Mach's principle
and supports Einstein's perspective on the physical universe.

We have reconstructed evolution histories of the closed world models
that are consistent with the recent astronomical observations, based on
the nearly flat FRW world models (Model I and II; Sec. \ref{sec:some}).
The present curvature radius of the universe is $a_0 = 25.7$ Gpc
($a_0=40.2$ Gpc) for Model I (Model II).
The expansion histories of both models imply that the closed universe
dominated by dark energy expands eternally.
However, the currently favored flat FRW world exists only as
a limiting case of the closed universe with infinite curvature radius
that is expanding with the maximum speed ($\dot{a}_0=1$, $\ddot{a}_0=0$).

From the local nature of the FRW metric (Sec. \ref{sec:metric})
and of the proper time of a comoving observer (Sec. \ref{sec:rel_friedmann}),
it is clear that the FRW world model
describes the local universe as observed by the comoving observer.
Since the Newton's gravitation law can be derived from the Einstein's field
equations in the weak field and the small velocity limits,
the gravitational action at a distance usually holds at a local region
of space on scales far smaller than the Hubble horizon size
(e.g., \cite{peebles80}).
The proper Hubble radius $d_\textrm{H}$ (Fig. \ref{fig:horizon}, top)
may provide a reasonable estimate of the characteristic distance scale
where the Newton's gravity applies. The cosmic structures simulated
by the Newton's gravity-based $N$-body method will significantly deviate
from the real structures on scales comparable to $d_\textrm{H}$.
The variation of the comoving Hubble radius
$\chi_\textrm{H} = d_\textrm{H} /a$ also implies
that in the past (future) the Newtonian dynamics was (will be)
applicable on smaller region of space compared to the size of the universe
(Fig. \ref{fig:horizon}, bottom).

In this paper, the history of the universe has been tentatively
reconstructed based on cosmological parameters of non-flat FRW world models.
The more general cosmological perturbation theory and parameter estimation
are essential for accurate reconstruction of the cosmic history.

\begin{acknowledgments}
This work has been supported by the Astrophysical Research Center
for the Structure and Evolution of the Cosmos (ARCSEC)
funded by the Korea Science and Engineering Foundation.
I thank Professor Cheongho Han in Chungbuk National
University for providing me with a research position in ARCSEC.
\end{acknowledgments}

\def\and{{and }}

\end{document}